\documentclass[%
 reprint,
 amsmath,amssymb,
 aps,
]{revtex4-2}

\usepackage{graphicx}
\usepackage{color}
\usepackage{dcolumn}
\usepackage{bm}

\begin{document}

\preprint{APS/123-QED}

\title{Kinetic energy density functional based on electron distribution\\ 
  on the energy coordinate to describe covalent bond}

\author{Hideaki Takahashi}
\email{ hideaki.takahashi.c4@tohoku.ac.jp}
\affiliation{Department of Chemistry, Graduate School of Science, Tohoku University,
 Sendai, Miyagi 980-8578, Japan}
 
\date{\today}


\begin{abstract}
The development of kinetic energy functional (KEF) is known as one of the most difficult subjects in the electronic density functional theory (DFT). In particular, the sound description of chemical bonds using a KEF is a matter of great significance  in the field of theoretical physics and chemistry. It can be readily confirmed that the famous Thomas-Fermi (TF) model or the TF model corrected with a generalized gradient approximation (GGA) fails to realize the bound state of a covalent bond in general. In this work, a new kinetic energy functional is developed on the basis of the novel density functional theory (J. Phys. B: At. Mol. Opt. Phys. \textbf{51}, 055102, 2018) that utilizes the electron distribution on the energy coordinate as the fundamental variable.  It is demonstrated for an H$_2$ molecule that the bound state can be realized by the KEF by virtue of the property of the electron density on the energy coordinate. The mechanism underlying the formation of the bound state is the same as that for the realization of the static correlation in the exchange energy described with the new DFT. We also developed a method termed potential gradient method to make a correction to the TF model instead of the GGA approach.   \\
\end{abstract}

\maketitle


%
%
%
%
%

\section{Introduction}
In the modern theoretical physics and chemistry, the density functional theory (DFT)\cite{rf:parr_yang_eng} for electrons provides an indispensable theoretical framework for studying the electronic properties of materials and molecules. It is based on the fact that the properties of a material in its electronic ground state can in principle be determined solely from its electron density $n$. However, it should be noted that in practice only the Kohn-Sham density functional theory (KS-DFT)\cite{hohenberg1964, rf:kohn1965pr} allows us to provide an efficient and practical theoretical tool for describing the electronic structure of the system. The most important characteristic of the KS-DFT is that the kinetic energy of the electrons in the system is being constructed in terms of the set of one-electron wave functions $\{\varphi_i\}$ instead of the density $n$. The introduction of the wave functions is being necessitated because there has been no kinetic energy functional (KEF) $E_{\rm{kin}}[n]$ that provides kinetic energy with sufficient accuracy. However, the use of one-electron wave functions $\{\varphi_i\}$ gives rise to an increase in the computational cost associated with imposing the orthonormalization conditions between the wave functions. The cost scales as $O(N^2)$ with respect to the number $N$ of electrons in the system, which becomes an obstacle for the KS-DFT to apply it to larger systems.  Provided that the total electronic energy of the system under consideration can be constructed by an explicit functional of the electron density $n$, the computational cost can be drastically reduced since the orthonormalization process is no longer needed. Such an approach is often referred to as orbital-free DFT (OF-DFT).\cite{Wesolowski2013}  Thus, in the field of computational materials science, the development of an accurate KEF is a matter of great significance to extend the application of the DFT to the systems with realistic sizes.\cite{rf:hung2010cpc}    \\
\indent  It is well known that the history of the KEF begun with the Thomas-Fermi (TF) model\cite{Thomas1927PCPS, Fermi1928zp} based on the homogeneous electron gas (HEG)\cite{rf:parr_yang_eng}. Within the TF or the Thomas-Fermi-Dirac (TFD) model, the electron density $n$ can be explicitly described by a \textit{local} function of the electrostatic potential $\Phi$ formed by the density $n$ and the nuclear charges. As a destructive conclusion, however, Bal\`{a}zs proved in general that no binding force can be created between the atoms in a molecule when the density $n$ is given by a function of $\Phi$.\cite{Balazs1967pr, rf:parr_yang_eng} To overcome the critical flaw, several approaches have been developed to improve the TF and TFD models. The gradient expansion approximation (GEA)\cite{Hodges1973CanJChem} provides a straightforward extension of the TF model, where the second and fourth derivatives of the density are included in the KEF to take into account for the effect of the inhomogeneity of the density under consideration. Unfortunately, it is known that the GEA approach offers only a minor improvement.\cite{Perdew1988prb} A series of the developments categorized in the generalized gradient approximation (GGA)\cite{Lee1991pra, Tran2002ijqc} uses the density $n$ and its gradient $\nabla n$ as the arguments, which shows rather better performances in computing the kinetic energies of atoms and molecules.\cite{Thakkar1992}  However, to the best of our knowledge, the benchmark test calculations for the GGA kinetic energy functional were performed only for atoms or molecules at their equilibrium structures.\cite{Lee1991pra, Thakkar1992, Tran2002ijqc} Other approaches by Vitos \textit{et al}\cite{rf:vitos2000pra} and Constantin \textit{et al}\cite{Constantin2009prb} are based on the inhomogeneous electron gas on the edge state\cite{rf:kohn1998prl} characterized by the slope $F$ of the local effective potential $\upsilon_{\rm{eff}}$ at the edge. The slope $F$ at the electron coordinate $\bm{r}$ can be related to the scaled gradient of the electron density.\cite{rf:vitos2000pra} Then, the kinetic energy density $\tau(\bm{r})$ of a real electron density with a certain scaled density can be described in terms of the kinetic energy density of the corresponding edge electron gas. In Ref. \cite{Constantin2009prb}, the potential energy curve of an \rm{N}$_2$ molecule was provided by using the given spin-unrestricted Hartree-Fock(HF) electron density.     

Another category of KEFs was provided by Wang and Teter (WT)\cite{rf:Wang1992prb}, who introduced a \textit{nonlocal} term in the functional to realize the formation of the shell structures in the electron densities around the atomic nuclei. The kernel $\omega(n_0,|\bm{r}-\bm{r}^\prime|)$ in the nonlocal term is described in terms of the linear response function of the HEG with the density $n_0$. For the bulk system, the average electron density in the unit cell is taken as the density $n_0$. It was shown in Ref. \cite{rf:Wang1992prb} the oscillatory behaviors appear in the densities in the atoms they considered.  The WT functional has been extensively studied and considerable efforts have been made to improve the nonlocal term.\cite{Wang1998prb, Wang1999prb} In particular, Huang and Carter (HC) proposed to modulate the nonlocal term by introducing the scaled gradient for applying the functional to semiconductors.\cite{Huang2010prb}  As a challenging application of the HC functional, the functional was employed for the calculations of the dissociation energy curves of the homonuclear diatomic molecules.\cite{Xia2012jcp}. It is shown in Ref. \cite{Xia2012jcp} that the bound states can be formed in the molecules computed using the OF-DFT with the HC functional. It seems, however, that the average electron density $n_0$ cannot be well defined for a molecule. Motivated by the WT functional, HT provided in Ref.\cite{Takahashi2022ijqc} a nonlocal kinetic functional that uses the composite linear response function (LRF) built from the LRFs of the constituent atoms in a hydrogen molecule. The important characteristic of the nonlocal functional is that the LRF defined on the real space is projected onto the energy coordinate $\epsilon$ on the basis of the framework of a novel  DFT\cite{rf:Takahashi2018, Takahashi2020jpb} where the energy electron density $n^e(\epsilon)$ serves as a fundamental variable. It is shown in Ref.\cite{Takahashi2022ijqc}, the potential energy curve of an H$_2$ molecule around its equilibrium bond length can be reasonably described with the nonlocal functional.  Hereafter, we use the superscript $e$ to emphasize that the functional is built for the DFT using the density $n^e(\epsilon)$ by virtue of the properties of the distribution $n^e(\epsilon)$. 

The aim of the present work is to develop a kinetic energy functional $E_{\rm{kin}}^e[n^e]$ which allows to describe dissociation of the covalent bond. It has often been the case that the performance of a KE functional $E_{\rm{kin}}[n]$ is to be evaluated by calculating the kinetic energy itself of a molecule at its equilibrium structure or the energy difference $\Delta E_{\rm{kin}}$ between the kinetic energies of the molecule and the isolated fragment atoms.\cite{Perdew1988prb, Thakkar1992, Constantin2009prb} However, it should be stressed that the description of a covalent bond with the OF-DFT has its own difficulties, as will be discussed later in more details. In fact, the source of the difficulties is essentially the same as the problem that arises in describing the static correlation associated with the bond dissociation using an exchange-correlation functional. The exchange hole model used in the local density approximation (LDA) is based on the HEG. LDA is appropriate for the isolated atom or molecule since the exchange hole is mostly localized in such a system. On the other hand, in the dissociation limit, the exchange hole is completely \textit{delocalized} over multiple sites, hence the hole model of the HEG does not work properly. As a consequence, the potential energy given by the symmetry adapted solution of the KS-DFT overshoots significantly the correct dissociation energy. Importantly, such an error occurs even in the dissociation of the simplest molecule, i.e. an H$_2$ molecule. Although the kinetic energy of the electrons has no direct relevance to the exchange hole function, the problem of evaluating the kinetic energy for the dissociating covalent bond using the TF model  is the same from a numerical point of view as that of the exchange energy computed with the LDA.  In Refs.\cite{rf:Takahashi2018, Takahashi2020jpb}, it was shown that the static correlation can be naturally realized by means of a simple exchange-correlation functional $E_{xc}^e[n^e]$. In the present work, we formulate the kinetic energy functional $E_{\rm{kin}}^e[n^e]$ in the way almost parallel to the development of the functional $E_{xc}^e[n^e]$. It will be shown that the kinetic energy functional $E_{\rm{kin}}^e[n^e]$ in combination with $E_{xc}^e[n^e]$ allows to correctly describe the dissociation of H$_2$ for the symmetry adapted electron density.  

The organization of the paper is as follows. In the next section, we will first discuss the source of the difficulty in describing the kinetic energy of the dissociating bond with a functional based on the TF model. Emphasis is placed on the numerical aspect of the problem in applying the TF model to the bond stretch. The source of the problem can then be directly compared to that of the static correlation error associated with the LDA applied to the bond dissociation. It will be shown that the problem in the TF model can be solved by the kinetic energy functional based on the distribution $n^e(\epsilon)$ in the way parallel to the static correlation functional.\cite{rf:Takahashi2018, Takahashi2020jpb} 
An approach to improve the kinetic energy functional $E_{\rm{kin}}^e[n^e]$ will also be provided using the GGA corrections\cite{Lee1991pra, Tran2002ijqc} for the TF model. In addition to the GGA approach, we will offer a novel correction method that utilizes the gradient of the external potential $|\nabla \upsilon_{\rm{ext}}|$ rather than the density gradient $|\nabla n|$. In the third section, the potential energy curves (PECs) for the dissociating H$_2$ molecule will be computed using the KEFs developed in the present work. The PECs will be compared with those given by the KS-DFT to discuss the problems in the functional $E_{\rm{kin}}^e[n^e]$.  In the section Conclusion, we will make a remark on the perspective of the future development of the functional based on the variable $n^e(\epsilon)$.

\section{Theory and Method}
\subsection{Thomas-Fermi Model and Static Correlation Error}
\begin{figure}[h]
\centering
\scalebox{0.30}[0.30] {\includegraphics[trim=70 60 80 70,clip]{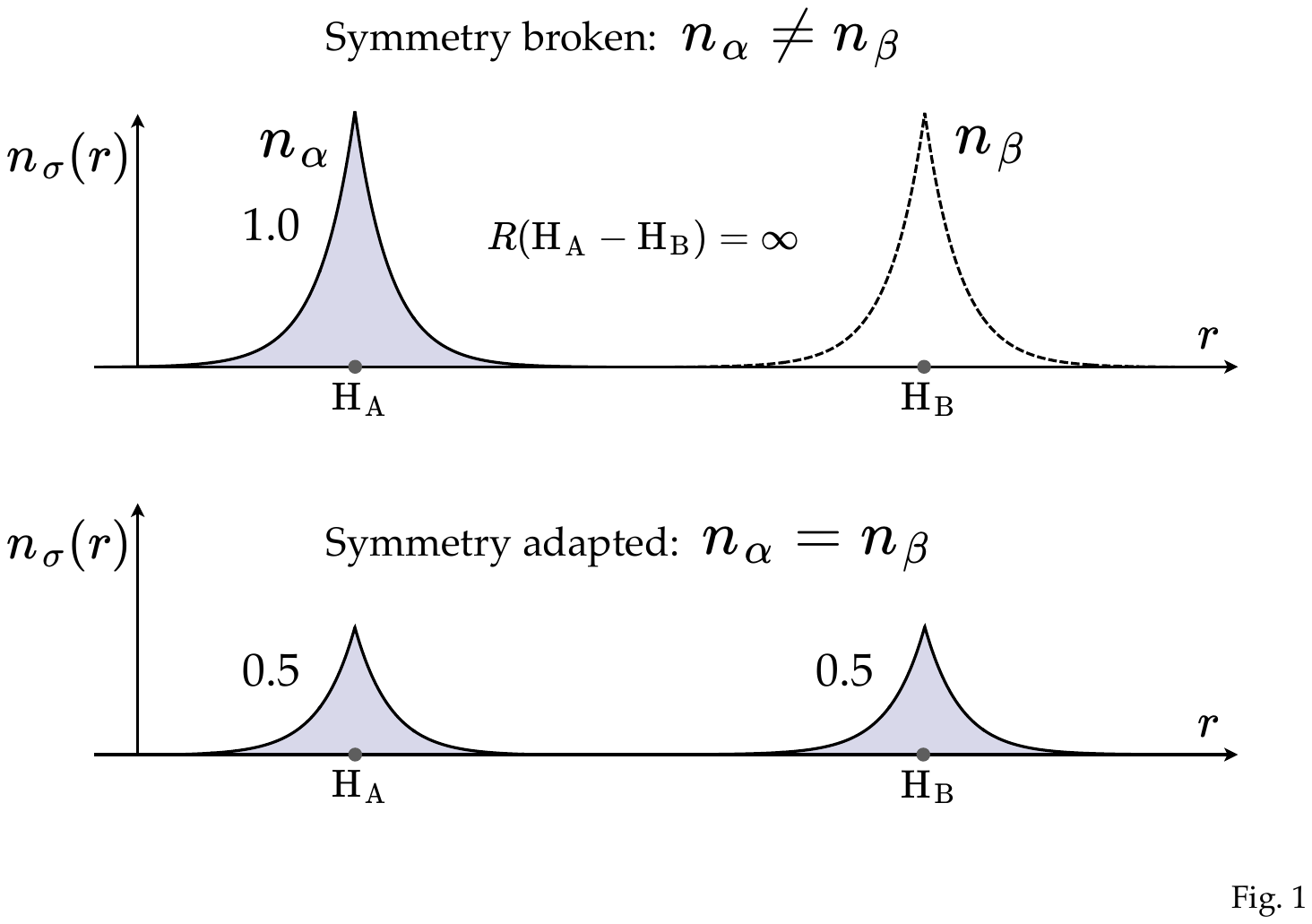}}      
\caption{\label{fig:TF_model}Electron densities $n_{\sigma} (\sigma = \alpha, \beta)$ of an H$_2$ molecule at the dissociation limit. Upper: symmetry broken electron density where an electron with spin $\alpha$ is localized at the left atom H$_{\rm{A}}$ and that with spin $\beta$ is at H$_{\rm{B}}$. Lower: symmetry adapted density where half an electron with spin $\sigma$ occupies each of the sites. }
\end{figure}

It is quite helpful to clarify the problem of the Thomas-Fermi (TF) model in describing the dissociation of a covalent bond for the later discussions. From a physical point of view, as mentioned above, Bal\`{a}zs already proved in general that no bound state can be formed between atoms by means of the TF or TFD model.\cite{Balazs1967pr, rf:parr_yang_eng} We consider here the numerical aspect of the problem associated with the models applied to the bond dissociations. Actually, the dissociation of an H$_2$ molecule is the simplest model sufficient to elucidate the mechanism underlying the problem. Figure \ref{fig:TF_model} shows the schematic illustrations of the electron densities of H$_2$ with the infinitely stretched bond. The upper figure represents the spin-polarized electron density, where an electron with spin $\alpha$ is located on the atom H$_{\rm{A}}$ and the electron with the opposite spin is on H$_{\rm{B}}$. On the other hand, in the lower figure, half an electron with spin $\alpha$ occupies each of the hydrogen atoms H$_{\rm{A}}$ and H$_{\rm{B}}$ and the same is true for the electron with the opposite spin. For the symmetry broken (SB) density $n^{\rm{SB}}$, the kinetic energy $E_{\rm{kin}}^{{\rm TF}}\left[n^{\rm{SB}}\right]$ by the TF model of the whole system can be described as
\begin{equation}     
E_{\rm{kin}}^{{\rm TF}}\left[n^{\rm{SB}}\right]=2^{2/3}\sum_{\sigma}C_{{\rm TF}}\int d\bm{r}n_{\sigma}\left(\bm{r}\right)^{\frac{5}{3}}
\label{eq:TF_symm_broken}
\end{equation}
where $n_{\sigma}(\bm{r})$ represents the electron density at the spatial coordinate $\bm{r}$ with spin $\sigma$ \textit{localized} at the atomic site A or B. It is supposed that $n_{\sigma}$ is normalized to $1.0$ and the coefficient $C_{\rm{TF}}$ in Eq. (\ref{eq:TF_symm_broken}) is given by $\frac{3}{10}(3\pi^2)^{2/3}$. On the other hand, in the symmetry adapted (SA) density $n^{\rm{SA}}$, the kinetic energy $E_{\rm{kin}}^{{\rm TF}}\left[n^{\rm{SA}}\right]$ is written as
\begin{equation}
E_{\rm{kin}}^{{\rm TF}}\left[n^{\rm SA}\right]=2\times2^{2/3}\sum_{\sigma}C_{{\rm TF}}\int d\bm{r}\left(\frac{1}{2}n_{\sigma}\left(\bm{r}\right)\right)^{\frac{5}{3}}
\label{eq:TF_symm_adapt}
\end{equation} 
where the multiplication by 2 implies the contributions from the densities on the atomic sites A and B. Then, it can be readily confirmed that the relation 
\begin{equation}
E_{\rm{kin}}^{{\rm TF}}\left[n^{\rm SA}\right]=0.629960\cdots \times E_{\rm{kin}}^{{\rm TF}}\left[n^{\rm SB}\right]
\label{eq:Ekin_nsb>Ekin_nsa}
\end{equation}
holds. Provided that the TF model gives a proper electronic kinetic energies individually for a hydrogen atom and the H$_2$ molecule, $E_{\rm{kin}}^{{\rm TF}}\left[n^{\rm{SB}}\right]$ delivers a reasonable kinetic energy change $\Delta E_{\rm{kin}}=E_{\rm{kin}}^{\rm{diss}}-E_{\rm{kin}}^{\rm{bond}}$ associated with the bond dissociation. However, $\Delta E_{\rm{kin}}$ will be seriously underestimated when the symmetry adapted density $n^{\rm{SA}}$ is used as the argument of the kinetic energy functional $E_{\rm{kin}}^{{\rm TF}}$ due to the direct consequence of Eq. (\ref{eq:Ekin_nsb>Ekin_nsa}).  
In other words, the TF energy will be erroneously lowered upon dissociation. Essentially the same discussion applies to the exchange energy change $\Delta E_{x}^{\rm{LDA}}$ evaluated with the local density approximation (LDA). In parallel to Eqs. (\ref{eq:TF_symm_broken}) and (\ref{eq:TF_symm_adapt}) we have the LDA exchange energies for the spin polarized and non-polarized densities, thus,
\begin{equation}     
E_{x}^{{\rm LDA}}\left[n^{\rm{SB}}\right]=-2^{1/3}\sum_{\sigma}C_{x}\int d\bm{r}n_{\sigma}\left(\bm{r}\right)^{\frac{4}{3}}
\label{eq:LDA_symm_broken}
\end{equation}
and  
\begin{equation}
E_{x}^{{\rm LDA}}\left[n^{\rm SA}\right]=-2\times2^{1/3}\sum_{\sigma}C_{x}\int d\bm{r}\left(\frac{1}{2}n_{\sigma}\left(\bm{r}\right)\right)^{\frac{4}{3}}
\label{eq:LDA_symm_adapt}
\end{equation} 
where the factor $C_x$ is given by $\frac{3}{4} \left(\frac{3}{\pi}\right)^{\frac{1}{3}}$. Then, we get
\begin{equation}
E_{x}^{{\rm LDA}}\left[n^{\rm SA}\right]=0.793701\cdots \times E_{x}^{{\rm LDA}}\left[n^{\rm SB}\right]
\label{eq:Ex_nsb>Ex_nsa}
\end{equation}
Thus, the absolute of the LDA exchange energy $E_{x}^{{\rm LDA}}\left[n^{\rm SA}\right]$ for the spin adapted density is seriously underestimated upon the bond dissociation, which leads to the overshoot of the dissociation energy of the H$_2$ molecule. This failure is well known as the static correlation error (SCE) in DFT. Unfortunately, it is also known that the error cannot be compensated by the GGA correction. Actually, it was shown in Ref. \cite{rf:becke2003jcp} that the SCE of a GGA functional becomes $+73$ mH at the dissociation limit of H$_2$. The SCE can be understood by the delocalization of the exchange hole function upon dissociation. Although the TF model is, of course, not relevant to the exchange hole, the source of the failure of the TF model in describing the bond dissociation is essentially the same as the SCE from the numerical point of view. Importantly, the situation is even worse for the kinetic energy of the TF model due to the fact that the energy scales as a power of $\frac{5}{3}$ of the density. Obviously, the error of the TF model cannot be corrected by any GGA corrections. One might think that the emergence of the error in the TF model as well as the SCE could be avoided by allowing the density to have a broken symmetry, although such electron configuration violates the principle of quantum mechanics. In fact, the SCE can be removed at least qualitatively by introducing the spin polarized density, where the degree of the polarization can be determined automatically from a self-consistent field (SCF) calculation based on the variational principle. In the TF model, however, the spin polarization \textit{increases} the total energy as well as the kinetic energy. Thus, the symmetry adapted density will be favored during the SCF calculation. Based on the above discussion, we propose below a method to overcome the difficulty by utilizing a new framework of DFT.\cite{rf:Takahashi2018, Takahashi2020jpb} 

\subsection{Construction of Thomas-Fermi Model on the Energy Coordinate}
The approach to the problem of the TF model in describing the dissociation curve of a covalent bond is actually parallel to the method developed to resolve the static correlation error (SCE).\cite{rf:Takahashi2018, Takahashi2020jpb}
Hence, the following discussion will be mostly parallel to the explanation of the exchange functional $E_{xc}$ which includes the static correlation. 
We use the hydrogen molecule illustrated in Fig. \ref{fig:TF_model}. Hereafter, in the present subsection, we omit the spin index $\sigma$ for the sake of notational simplicity.  
It is assumed that a reference density $n_0$ is provided for which the kinetic energy can be properly obtained. It is quite natural to take $n_0$ as the sum of the ground state densities $n_0^{\rm{A}}$ and $n_0^{\rm{B}}$ of the isolated atoms H$_{\rm{A}}$ and H$_{\rm{B}}$, thus, 
\begin{equation}
n_{0}\left(\bm{r}\right)=n_0^{\rm{A}}\left(\bm{r}\right)+n_0^{\rm{B}}\left(\bm{r}\right)
\label{eq:ref_density}
\end{equation}
Such a reference state has already been introduced in the work by Harris \cite{rf:harris1985prb}, and a similar approach can also be found in the context of the frozen density functional method\cite{Wesolowski1993jpc} or the partition DFT\cite{rf:elliott2010pra}, where the density of the entire system is provided by the overlap of the densities of the partitioned molecular regions.  

Anyway, in the present case, $n_0^{\rm{A}}$ is the density with spin $\alpha$ localized at H$_{\rm{A}}$ as shown in Fig. \ref{fig:TF_model} (upper left), while $n_0^{\rm{B}}$ is that with spin $\beta$ located at H$_{\rm{B}}$ (upper right). In the construction of the density of Eq. (\ref{eq:ref_density}), it is assumed that there exists no interaction between two hydrogen atoms H$_{\rm{A}}$ and H$_{\rm{B}}$ regardless of their nuclear distance $R$. Now consider a gradual introduction of the interaction between the atoms, which leads to the variation of the density from $n_0$ to $n_1$ with $n_1$ being the spin adapted ground state density (lower figure in Fig. \ref{fig:TF_model}) of a fully interacting system. By using the coupling parameter $\lambda$ that describes the gradual variation from $n_0$ to $n_1$, a kinetic energy functional $E_{\rm{kin}}[n_1]$ can be formally described as 
\begin{equation}
 E_{\rm{kin}}^{{\rm }}\left[n_{1}\right]=E_{\rm{kin}}\left[n_{0}\right]+\int_{0}^{1}d\lambda\;\frac{dE_{\rm{kin}}^{{\rm }}\left[n_{\lambda}\right]}{d\lambda}
\label{eq:coupling_int}
\end{equation}
Our strategy is to replace only the functional $E_{\rm{kin}}$ in the integrand in Eq. (\ref{eq:coupling_int}) by the functional $E_{\rm{kin}}^e$ for the energy electron density $n^e(\epsilon)$ defined on the energy coordinate $\epsilon$, which reads
\begin{eqnarray}
\overline{E}_{\rm{kin}}\left[n_{1}\right] & = E_{\rm{kin}}\left[n_{0}\right]+\int_{0}^{1}d\lambda\;\frac{dE_{\rm{kin}}^{e}\left[n_{\lambda}^{e}\right]}{d\lambda}   \nonumber   \\
 & = E_{\rm{kin}}\left[n_{0}\right]+\left(E_{\rm{kin}}^{e}\left[n_{1}^{e}\right]-E_{\rm{kin}}^{e}\left[n_{0}^{e}\right]\right)
\label{eq:Ekin_func}
\end{eqnarray}   
The energy electron density $n^e(\epsilon)$ is defined as the projection of the density $n(\bm{r})$ onto the energy coordinate $\epsilon$, thus, 
\begin{equation}
n^{e}\left(\epsilon\right)=\int d\bm{r}\;n\left(\bm{r}\right)\delta\left(\epsilon-\upsilon_{{\rm ext}}\left(\bm{r}\right)\right)
\label{eq:dnst_e}
\end{equation}
where $\upsilon_{\rm{ext}}(\bm{r})$ is the external potential of the system under consideration. In the present case, $\upsilon_{{\rm ext}}$ is explicitly expressed as
\begin{equation}
\upsilon_{{\rm ext}}\left(\bm{r}\right)=-\sum_{a}\frac{1}{\left|\bm{r}-\bm{R}_{a}\right|}
\label{eq:v_ext}
\end{equation}
where $a$ is the index to specify the nuclei H$_{\rm{A}}$ and H$_{\rm{B}}$, and $\bm{R}_a$ is the coordinate for the nucleus $a$. As discussed in Ref. \cite{rf:Takahashi2018}, it is also possible to formulate the DFT for $n^e(\epsilon)$ by virtue of the one-to-one correspondence between the subset of $n^e(\epsilon)$ and the subset of $\upsilon_{\rm{ext}}(\bm{r})$.  We refer the reader to Ref. \cite{rf:Takahashi2018} for more details. 
Note that Eq. (\ref{eq:Ekin_func}) is completely parallel to the $E_{xc}$ functional that includes the static correlation expressed by Eqs. (28) and (29) in Ref. \cite{rf:Takahashi2018}. Then, it is possible to define the \textit{kinetic} static correlation $E_{\rm{kin}}^{\rm{sc}}[n_0]$ in parallel to the definition of Eq. (30) in Ref. \cite{rf:Takahashi2018}. $E_{\rm{kin}}^{\rm{sc}}[n_0]$ can be expressed as 
\begin{equation}
E_{\rm{kin}}^{\rm{sc}}[n_0] = E_{\rm{kin}}\left[n_{0}\right]-E_{\rm{kin}}^{e}\left[n_{0}^{e}\right]
\label{eq:Ekin_sc}
\end{equation}
Of course, the term \textquoteleft static correlation\textquoteright\; has its origin in the multi-configuration character of the wave function and it has no direct relevance to the kinetic energy. The use of the term is based solely on the similarity of the formula of Eq. (\ref{eq:Ekin_sc}) to Eq. (30) in Ref. \cite{rf:Takahashi2018}. It is quite important to note that, at the dissociation limit of an H$_2$ molecule, the quantity $E_{\rm{kin}}^{e}\left[n_{1}^{e}\right]-E_{\rm{kin}}^{e}\left[n_{0}^{e}\right]$ in Eq. (\ref{eq:Ekin_func}) strictly vanishes in principle since $n_1^e = n_0^e$ is attained for each spin although $n_1$ completely differs from $n_0$ in the real space. Thus, only the kinetic energy $E_{\rm{kin}}\left[n_{0}\right]$ of the isolated atoms remains upon dissociation. Importantly, Eq. (\ref{eq:Ekin_func}) delivers the correct kinetic energy at the dissociation for any kind of the functional $E_{\rm{kin}}^e$,  provided that $E_{\rm{kin}}[n_0]$ is properly evaluated.  

For the numerical implementation of Eq. (\ref{eq:Ekin_func}), the TF model can be directly applied. Explicitly, $E_{\rm{kin}}^{e}[n^e]$ in Eq. (\ref{eq:Ekin_func}) can be expressed as 
\begin{equation}
E_{{\rm kin}}^{e,{\rm TF}}\left[n^{e}\right]=2\times2^{2/3}C_{{\rm TF}}\int d\epsilon\;{n}^{e}\left(\epsilon\right)\widetilde{n}^{e}\left(\epsilon\right)^{\frac{2}{3}}
\label{eq:TF_e}
\end{equation}
where the factor $2$ in the right hand side represents the contributions from $\alpha$ and $\beta$ spins. $\widetilde{n}^{e}\left(\epsilon\right)$ is the average electron density of the electrons on the energy coordinate $\epsilon$ and defined as 
\begin{equation}
\widetilde{n}^{e}\left(\epsilon\right)=\frac{n^{e}\left(\epsilon\right)}{\Omega^{e}\left(\epsilon\right)} 
\label{eq:tilde_dnst_e}
\end{equation}
where $\Omega^e(\epsilon)$ is the spatial volume with the energy coordinate $\epsilon$ and is given by
\begin{equation}
\Omega^{e}\left(\epsilon\right)=\int d\bm{r}\;\delta\left(\epsilon-\upsilon_{{\rm ext}}\left(\bm{r}\right)\right)
\end{equation}
Note that the structure of Eq. (\ref{eq:TF_e}) is parallel to that of Eq. (24) in Ref. \cite{rf:Takahashi2018} constructed for the LDA exchange functional, which also uses the energy electron density $n^e$ as an argument. 
It would be helpful to note the property of the TF functional in Eq. (\ref{eq:TF_e}) for later discussions. The functional $E_{{\rm kin}}^{e,{\rm TF}}\left[n^{e}\right]$ becomes identical to $E_{{\rm kin}}^{{\rm TF}}\left[n\right]$ in Eq. (\ref{eq:TF_symm_broken}) for the ground state density of a hydrogenic atom because the electron density $n(\bm{r})$ as well as the corresponding external potential $\upsilon_{\rm{ext}}(\bm{r})$ are spherically symmetric around the nucleus. Substituting Eq. (\ref{eq:TF_e}) into Eq. (\ref{eq:Ekin_func}), we obtain the kinetic energy functional $\overline{E}_{\rm{kin}}^{\rm{TF}}[n]$ with the static correlation, thus, 
\begin{equation}
\overline{E}_{{\rm kin}}^{{\rm TF}}\left[n_{1}\right]=E_{{\rm kin}}^{{\rm TF}}\left[n_{0}\right]+\left(E_{{\rm kin}}^{e,{\rm TF}}\left[n_{1}^{e}\right]-E_{{\rm kin}}^{e,{\rm TF}}\left[n_{0}^{e}\right]\right)
\label{eq:SC_TF}
\end{equation}     

It is also possible to apply a GGA correction to the TF model of Eq. (\ref{eq:TF_e}) in parallel to the method in Ref. \cite{rf:Takahashi2018}. Remind that the inhomogeneity parameter $s(\bm{r})$ for each spin plays a role in kinetic GGA functionals in general\cite{Lee1991pra, Tran2002ijqc}, which is defined as 
\begin{equation}
s(\bm{r})=\frac{\left|\nabla n\left(\bm{r}\right)\right|}{n\left(\bm{r}\right)^{\frac{4}{3}}}
\end{equation}
For the absolute of the gradient of the density $G_{1}\left[n\right]\left(\bm{r}\right)\equiv\left|\nabla n\left(\bm{r}\right)\right|$, we construct its average over the energy coordinate $\epsilon$,   
\begin{equation}
G_{1}^{e}\left[n\right]\left(\epsilon\right)=\Omega^{e}\left(\epsilon\right)^{-1}\int d\bm{r}\;G_{1}\left[n\right]\left(\bm{r}\right)\delta\left(\epsilon-\upsilon_{{\rm ext}}\left(\bm{r}\right)\right)
\end{equation}
Note that if we replace $G_1[n](\bm{r})$ by $G_0[n](\bm{r})\equiv n(\bm{r})$, we get $G_0^e[n](\epsilon)=\widetilde{n}^{e}\left(\epsilon\right)$ accordingly. 
Using the function $G_{1}^{e}\left[n\right]\left(\epsilon\right)$, the inhomogeneity parameter $s^e(\epsilon)$ on the energy coordinate can be given by
\begin{equation}
s^{e}\left(\epsilon\right)=\frac{G_{1}^{e}\left[n\right]\left(\epsilon\right)}{\widetilde{n}^{e}\left(\epsilon\right)^{\frac{4}{3}}}
\end{equation}
Then, the TF model with a GGA correction can be formally expressed as
\begin{align}
E_{{\rm kin}}^{e,{\rm TF+GGA}}\left[n^{e}\right]&=2\times2^{2/3}C_{{\rm TF}}    \notag     \\
\times&\int d\epsilon\;n^{e}\left(\epsilon\right)\widetilde{n}^{e}\left(\epsilon\right)^{\frac{2}{3}}F_{{\rm kin}}^{e,{\rm GGA}}\left(s^{e}\left(\epsilon\right)\right)
\label{eq:TF+GGA}
\end{align}
where function $F_{{\rm kin}}^{e,{\rm GGA}}$ is often referred to as the enhancement factor. 
In the present implementation, we employ the enhancement factor $F_{\rm kin}^{\rm TW}(s)$ developed by Tran and Weso\l owski (TW) in Ref. \cite{Tran2002ijqc}. The formula of $F_{\rm kin}^{\rm TW}$ is common to that of the exchange GGA functional by Perdew \textit{et al}\cite{rf:perdew1996prl} and its explicit form defined on the energy coordinate can be written as
\begin{equation}
F_{{\rm kin}}^{e,{\rm TW}}\left(s^{e}\right)=1+\frac{\mu\cdot\left(s^{e}\right)^{2}}{1+\frac{\mu}{\kappa}\left(s^{e}\right)^{2}}
\label{eq:enhcf_e}
\end{equation}
where the fitting parameters $\kappa$ and $\mu$ were set at $\kappa = 0.8438$ and $\mu = 0.2319$, respectively\cite{Tran2002ijqc} for the kinetic energy functional. The kinetic GGA functional of Eq. (\ref{eq:TF+GGA}) defined for the distribution $\widetilde{n}^e(\epsilon)$ becomes equivalent to the corresponding functional defined for $n(\bm{r})$ when the functionals are applied to a hydrogenic atom since the density gradient $\left|\nabla n\left(\bm{r}\right)\right|$ is constant over a surface with the same energy coordinate $\epsilon$. We apply the kinetic GGA correction only to the coupling-parameter integration in Eq. (\ref{eq:Ekin_func}) because it was found in the preliminary calculations that applying the GGA correction also to the term $E_{\rm{kin}}[n_0]$ gives rise to an erroneous upward shift of the potential energy curve by $\sim0.1$ au. The overestimation of the GGA correction is due to the fact that kinetic static correlation $E_{\rm{kin}}^{\rm{sc}}[n_0]$ defined in Eq. (\ref{eq:Ekin_sc}) itself includes the energy corresponding to the GGA correction. Anyway, the GGA correction to $E_{\rm{kin}}[n_0]$ is constant regardless of the distance $R$ between the nuclei, so the  potential profile is not affected by the correction. In summary, a GGA corrected kinetic energy functional is written as
\begin{align}     
\overline{E}_{\rm{kin}}^{\rm{TF+GGA}}\left[n_{1}\right] &=E_{\rm{kin}}^{{\rm TF}}\left[n_{0}\right]   \notag   \\
& +\left(E_{\rm{kin}}^{e,{\rm TF+GGA}}\left[n_{1}^{e}\right]-E_{\rm{kin}}^{e,{\rm TF+GGA}}\left[n_{0}^{e}\right]\right)  
\end{align}
In the present work, the TW functional is used as the GGA correction. Specifically, it is written as
\begin{align}     
\overline{E}_{\rm{kin}}^{\rm{TF+TW}}\left[n_{1}\right]&=E_{\rm{kin}}^{{\rm TF}}\left[n_{0}\right]    \notag  \\
&+\left(E_{\rm{kin}}^{e,{\rm TF+TW}}\left[n_{1}^{e}\right]-E_{\rm{kin}}^{e,{\rm TF+TW}}\left[n_{0}^{e}\right]\right)
\label{eq:SC_TF_TW}
\end{align} 
In the next subsection we also provide a new method to correct the TF energy by using the gradient of the external potential instead of the density gradient.  

\subsection{Potential Gradient Approximation}
In the generalized gradient approximation (GGA), the absolute of the density gradient plays a fundamental role to correct the Thomas-Fermi (TF) model. The density gradient is, of course, directly induced by the slope of the  potential that acts on the electrons. It is, thus, natural to employ the potential gradient $q(\bm{r}) \equiv \left| \nabla \upsilon_{\rm{ext}}(\bm{r})\right|$ instead of $\left| \nabla n(\bm{r})\right|$ to correct the TF energy. In fact, the increase in the slope of the potential gives rise to the monotonic increase in the enhancement factor as proved for a hydrogenic atom. The average potential gradient $q^e(\epsilon)$ on the energy coordinate $\epsilon$ can be given by
\begin{equation}
q^{e}\left(\epsilon\right)=\Omega^{e}\left(\epsilon\right)^{-1}\int d\bm{r}\;\left|\nabla\upsilon_{{\rm ext}}\left(\bm{r}\right)\right|\delta\left(\epsilon-\upsilon_{{\rm ext}}\left(\bm{r}\right)\right)
\label{eq:PG_e}
\end{equation}   
Then, the potential gradient approximation (PGA) to correct the TF model is expressed as
\begin{align}
E_{{\rm kin}}^{e,{\rm TF+PGA}}\left[n^{e}\right]&=2\times2^{2/3}C_{{\rm TF}}     \notag      \\
\times&\int d\epsilon\;n^{e}\left(\epsilon\right)\widetilde{n}^{e}\left(\epsilon\right)^{\frac{2}{3}}Q_{{\rm kin}}^{e,{\rm PGA}}\left(q^{e}\left(\epsilon\right)\right)
\end{align}
In the present work, the enhancement factor $Q_{{\rm kin}}^{e,{\rm PGA}}$ is given by
\begin{equation}
Q_{{\rm kin}}^{e,{\rm PGA}}\left(q^{e}\right)=1+\frac{\beta\cdot\left(q^{e}\right)^{2}}{1+\gamma\cdot\left(q^{e}\right)^{2}}
\label{eq:PGA_enhcf_e}
\end{equation} 
where $\beta$ and $\gamma$ are the fitting parameters. Note the complete parallelism of the formula to the enhancement factor in the GGA functional of Eq. (\ref{eq:enhcf_e}). Then, the kinetic energy functional $\overline{E}_{\rm{kin}}^{\rm{TF+PGA}}\left[n_{1}\right]$ with the PGA correction is written as
\begin{align}     
\overline{E}_{\rm{kin}}^{\rm{TF+PGA}}\left[n_{1}\right]&=E_{\rm{kin}}^{{\rm TF}}\left[n_{0}\right]     \notag   \\
&+\left(E_{\rm{kin}}^{e,{\rm TF+PGA}}\left[n_{1}^{e}\right]-E_{\rm{kin}}^{e,{\rm TF+PGA}}\left[n_{0}^{e}\right]\right)
\label{eq:SC_TF_PGA}
\end{align} 
in parallel to Eq. (\ref{eq:SC_TF_TW}). 
Since the major purpose of this work is not to optimize the parameters but to examine the efficiency of the PGA method, the fitting parameters in Eq. (\ref{eq:PGA_enhcf_e}) are roughly determined as $\beta = \gamma = 4.0\times10^{-3}$. 

\subsection{Numerical Details}
For the numerical implementation of the kinetic energy functionals (KEFs) described in Eqs. (\ref{eq:SC_TF_TW}) and (\ref{eq:SC_TF_PGA}), a program module has been added to the \textquoteleft Vmol\textquoteright\; code developed for the real-space grid KS-DFT.\cite{rf:takahashi2000cl, takahashi2001jpca, rf:takahashi2001jcc, rf:takahashi2003jcp} The electron density $n_1$ and the density $n_0$ for the reference system are obtained through the self-consistent field (SCF) calculation of the KS-DFT.  In the real-space grid approach\cite{rf:chelikowsky1994prl, rf:chelikowsky1994prb}, the Coulomb interactions between the electrons and the nuclei are described using nonlocal pseudopotentials $\hat{\upsilon}_{\rm{ps}}$. Then, the KS equation for the real-space grid method is expressed as
\begin{eqnarray}
\left(-\frac{1}{2}\nabla^{2}+\upsilon_{{\rm H}}\left[n\right]\left(\bm{r}\right)  +\upsilon_{xc}\left[n\right]\left(\bm{r}\right)\right)  \varphi_{i}\left(\bm{r}\right)  \nonumber   \\
\;\;\;\;\;\;\;\;\;\;\;\;\; +\int d\bm{r}^{\prime}\upsilon_{{\rm ps}}\left(\bm{r},\bm{r}^{\prime}\right)\varphi_{i}\left(\bm{r}^{\prime}\right) =\varepsilon_{i}\varphi_{i}\left(\bm{r}\right)
\label{eq:KSeq}
\end{eqnarray}
where $\{\varphi_i(\bm{r})\}$ and $\{\varepsilon_i\}$ are the KS wave functions and the corresponding eigenvalues for each spin, respectively. The density $n$ is given by $n\left(\bm{r}\right)=2\sum_{i}\left|\varphi_{i}\left(\bm{r}\right)\right|^{2}$. $\upsilon_{\rm{H}}[n](\bm{r})$ in Eq. (\ref{eq:KSeq}) is the Hartree potential defined by
\begin{equation}
\upsilon_{{\rm H}}\left[n\right]=\int d\bm{r}^{\prime}\frac{n\left(\bm{r}^{\prime}\right)}{\left|\bm{r}-\bm{r}^{\prime}\right|}\;\;\;, 
\end{equation}
which is computed by the method proposed in Ref. \cite{Barnett1993prb}. $\upsilon_{xc}[n](\bm{r})$ is the exchange-correlation potential given by the functional derivative of the exchange-correlation functional $E_{xc}[n]$ with respect to $n(\bm{r})$, that is $\upsilon_{xc}[n](\bm{r})=\frac{\delta E_{xc}[n]}{\delta n(\bm{r})}$.  

The Becke88 exchange functional\cite{rf:becke1988pra} combined with the LYP correlation energy\cite{rf:lee1988prb}, abbreviated as BLYP, is employed as the exchange-correlation functional $E_{xc}[n]$. For the hydrogen atom, the nonlocal pseudopotential $\hat{\upsilon}_{\rm{ps}}$ in the form proposed by Kleinman and Bylander\cite{rf:kleinman1982prl} is adopted. The double grid technique\cite{rf:ono1999prl} is used to realize the steep behavior of the pseudopotential near the atomic core region. The 4th-order finite difference method\cite{rf:chelikowsky1994prl, rf:chelikowsky1994prb} is applied to represent the kinetic energy operator in Eq. (\ref{eq:KSeq}). A cubic cell with size $L=18.354$ a.u. is employed to enclose the wave function of the H$_2$ molecule. Each axis of the cell is uniformly discretized by 64 grids, which leads to the grid width $h=0.28679$ a.u. The size $d$ of the double grid is set at $d=h/5$. 

The energy electron density $n^e(\epsilon)$ defined in Eq. (\ref{eq:dnst_e}) is constructed on the energy coordinate $\epsilon$ discretized by a natural logarithmic grid $\epsilon_m$ specified by
\begin{equation}
\log\epsilon_{m}=\frac{m-1}{M}\log\left(\frac{\epsilon_{{\rm max}}}{\epsilon_{{\rm min}}}\right)+\log\epsilon_{{\rm min}}
\end{equation}     
where $M$ is the number of grid points and $\epsilon_{{\rm min}}\leq\epsilon_{m}\leq\epsilon_{{\rm max}}$ defines the range of the energy coordinate. In defining the energy coordinate $\epsilon$ with Eq. (\ref{eq:v_ext}), we use the potential with the opposite sign. This treatment reverses only the direction of the coordinate $\epsilon$ without changing the physical contents. In the present implementation, we adopt the settings; $M=20$, $\epsilon_{\rm{min}}=0.1$ a.u. and $\epsilon_{\rm{max}}=7.0$ a.u. To avoid the divergence of the energy coordinate at the atomic cores, the absolute of the local part in the pseudopotential developed by Bachelet, Hamann, and Schl\"{u}ter (BHS) \cite{Bachelet1982prb} for the hydrogen atom is employed to define the energy coordinate instead of the bare Coulomb potential. To increase the number of sampling points for constructing $n^e(\epsilon)$, each grid in the real-space cell is discretized by $5$ points along each axis and the density $n(\bm{r})$ on the dense grid is evaluated through the 4th-order polynomial interpolation of the coarse grids. 

The total electronic energy of an H$_2$ molecule, that includes the kinetic energy $\overline{E}_{{\rm kin}}^{{\rm A}}\left[n\right]$, is evaluated by the following formula,
\begin{align}
E_{\rm tot}^{\rm{A}}\left[n, \{\varphi_i\}\right] & = \overline{E}_{{\rm kin}}^{{\rm A}}\left[n\right]+\overline{E}_{x}^{{\rm LDA}}\left[n\right]   \notag  \\
 & \; +E_{{\rm BLYP}}^{\prime}\left[n\right]+E_{{\rm H}}\left[n\right] + E_{{\rm ext}}\left[\{\varphi_i\}\right]
\label{eq:E_total}
\end{align}
where the superscript \textquoteleft A\textquoteright\; stands for the kinetic energy functionals \textquoteleft TF\textquoteright , \textquoteleft TF+TW\textquoteright\; and \textquoteleft TF+PGA\textquoteright , specified by Eqs. (\ref{eq:SC_TF}), (\ref{eq:SC_TF_TW}) and (\ref{eq:SC_TF_PGA}), respectively.  In Eq. (\ref{eq:E_total}), the term $\overline{E}_{x}^{{\rm LDA}}\left[n\right]$ represents the LDA exchange energy including the static correlation\cite{rf:Takahashi2018} and is given by 
\begin{equation}
\overline{E}_{x}^{{\rm LDA}}\left[n\right]=E_{x}^{{\rm LDA}}\left[n_{0}\right]+\left(E_{x}^{e,{\rm LDA}}\left[n^{e}\right]-E_{x}^{e,{\rm LDA}}\left[n_{0}^{e}\right]\right)
\label{eq:Ex_sc}
\end{equation}
in parallel to Eq. (\ref{eq:Ekin_func}). $E_{x}^{{\rm LDA}}\left[n\right]$ in the right hand side of Eq. (\ref{eq:Ex_sc}) is the normal LDA exchange energy, while $E_{x}^{e,{\rm LDA}}\left[n^{e}\right]$ is that evaluated using the energy electron density $n^e(\epsilon)$. Explicitly, it is written as
\begin{equation}
E_{x}^{e,{\rm LDA}}\left[n^{e}\right]= - 2\times2^{1/3}C_{x}\int d\epsilon\;n^{e}\left(\epsilon\right)\widetilde{n}^{e}\left(\epsilon\right)^{\frac{1}{3}}
\end{equation}
with $C_{x}=\frac{3}{4}\left(\frac{3}{\pi}\right)^{1/3}$. Remind that the reference density $n_0(\bm{r})$ and $n_0^e(\epsilon)$ are included in the functionals $\overline{E}_{{\rm kin}}^{{\rm TF+TW}}\left[n\right]$ and $\overline{E}_{x}^{{\rm LDA}}\left[n\right]$ as shown in Eqs. (\ref{eq:SC_TF_TW}) and (\ref{eq:Ex_sc}). $n_0$ is provided by the overlap of the electron densities for the isolated hydrogen atoms as written in Eq. (\ref{eq:ref_density}).  

$E_{{\rm BLYP}}^{\prime}\left[n\right]$ in Eq. (\ref{eq:Ex_sc}) is given by the sum of the GGA correction term in the Becke88 functional and the LYP correlation energy for the normal density $n(\bm{r})$. Thus, the LDA energy $E_x^{\rm{LDA}}[n]$ is \textit{not} contained in $E_{{\rm BLYP}}^{\prime}\left[n\right]$. The last term in the right hand side of Eq. (\ref{eq:E_total}) is the Coulomb interaction between the electrons and the external potential due to the nuclei. Since no effective \textit{local} pseudopotential is available for a hydrogen atom, we employ a nonlocal pseudopotential and the KS wave functions $\{\varphi_i\}$ obtained as the solution of Eq. (\ref{eq:KSeq}). Specifically for the Kleinman-Bylander (KB) potential $\hat{\upsilon}_{\rm{ps}}^{\rm{KB}}$\cite{rf:kleinman1982prl}, $E_{{\rm ext}}\left[\{\varphi_i\}\right]$ is calculated as  
\begin{equation}  
E_{{\rm ext}}\left[\left\{ \varphi_{i}\right\} \right]=2\sum_{i}\int d\bm{r}d\bm{r}^{\prime}\varphi_{i}^{*}\left(\bm{r}\right)\upsilon_{{\rm ps}}^{\rm{KB}}\left(\bm{r},\bm{r}^{\prime}\right)\varphi_{i}\left(\bm{r}^{\prime}\right)
\end{equation}
Thus, the total electronic energy $E_{{\rm tot}}\left[n, \{\varphi_i\}\right]$ necessitates the KS wave functions. However, it does not matter since the major purpose of the present work is to examine the properties of the kinetic energy functionals of Eqs. (\ref{eq:SC_TF_TW}) and (\ref{eq:SC_TF_PGA}) for a given electron density. 

\section{Results and Discussion}
\subsection{Energy Electron Density}
\begin{figure}[h]
\centering
\scalebox{0.3}[0.3] {\includegraphics[trim=10 100 0 50,clip]{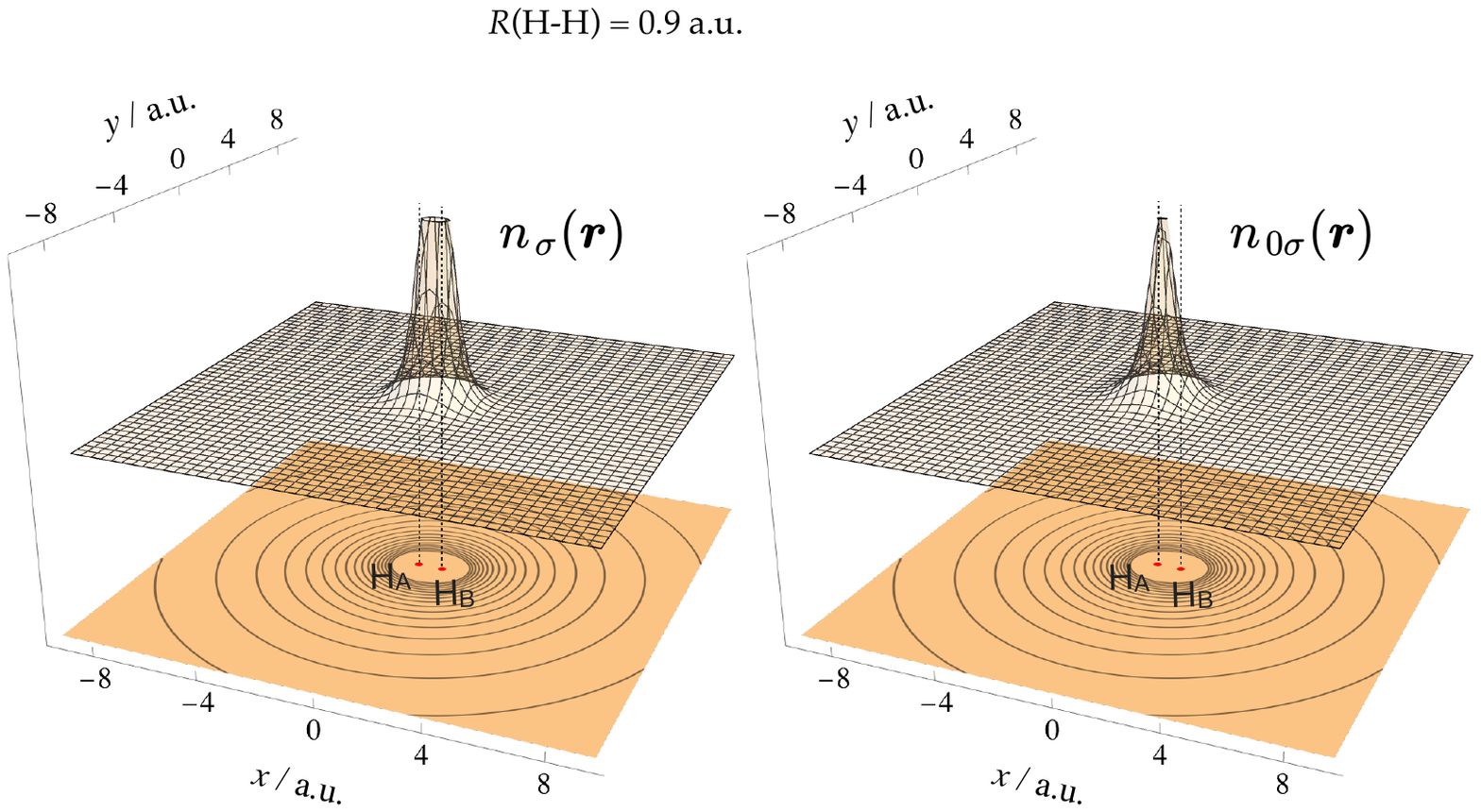}}      
\caption{\label{fig:density_09au} Left: 3-dimensional representation of the ground state electron density $n_{\sigma}(\bm{r}) (\sigma = \alpha\; \rm{or}\; \beta)$ of an H$_2$ molecule at $R$(H-H)$\;=0.9$ a.u. Right: Reference electron density $n_{0{\sigma}}(\bm{r})$ for the same system. The contour map at the bottom of each graph shows the absolute of the external potential $|\upsilon_{\rm{ext}}(\bm{r})|$ of the molecule. For these plots, the logarithms of the values are taken. Explicitly, the value $\log(Q+1.0)\; (Q = n \; \rm{or}\; |\upsilon_{\rm{ext}}|)$ are plotted. The plot range is $0\leq \log (n+1.0)\leq0.2$ for the density, and $0\leq \log (|\upsilon_{\rm{ext}}|+1.0)\leq 1.0$ for the potential. The BLYP functional\cite{rf:becke1988pra, rf:lee1988prb} is utilized for the construction of the densities. The local part of the BHS pseudopotential\cite{Bachelet1982prb} is used to describe the contour plot of the potential. }
\end{figure}

\begin{figure}[h]
\centering
\scalebox{0.3}[0.3] {\includegraphics[trim=10 100 0 50,clip]{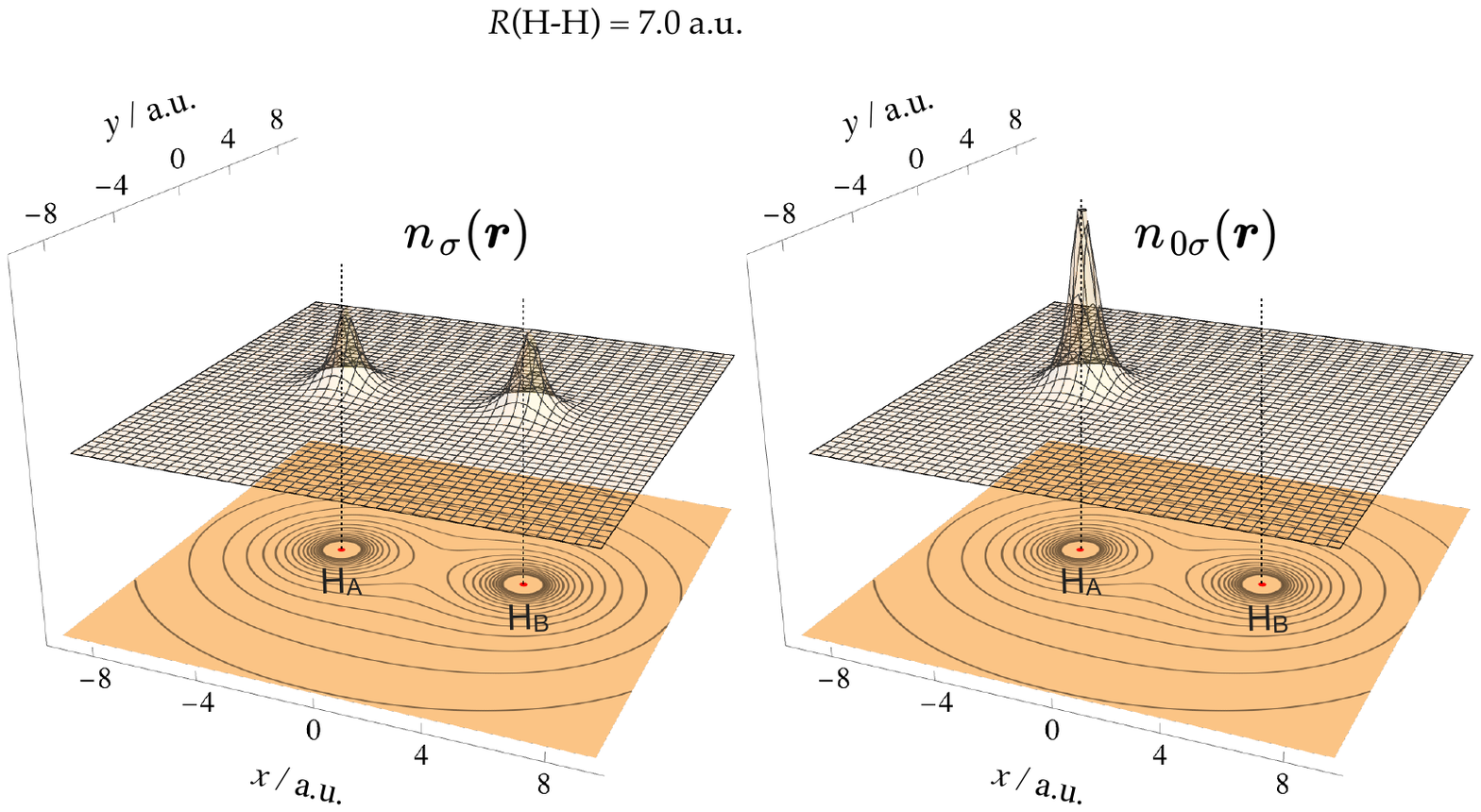}}      
\caption{\label{fig:density_7au}Left: 3-dimensional representation of the ground state electron density $n_{\sigma}(\bm{r}) (\sigma = \alpha\; \rm{or}\; \beta)$ of an H$_2$ molecule at $R$(H-H)$\;=7.0$ a.u. Right: Reference electron density $n_{0{\sigma}}(\bm{r})$ for the same system. The other notations are the same as in Fig. \ref{fig:density_09au}. }
\end{figure}

First, we provide graphical representations of the electron densities $n(\bm{r})$ and $n_0(\bm{r})$ for H$_2$ molecules with $R$(H-H)$\;=0.9$ and $7.0$ a.u., respectively, in Figs. \ref{fig:density_09au} and \ref{fig:density_7au}. It is shown in Fig. \ref{fig:density_09au} that the center of the reference density $n_{0\sigma}(\bm{r})$ is placed at the hydrogen atom H$_{\rm{A}}$. It is then assumed that the density $n_0(\bm{r})$ with the opposite spin is placed at H$_{\rm{B}}$ according to Eq. (\ref{eq:ref_density}). Note that the electron distribution of each spin is to be optimized for a hydrogen atom at an isolation. Due to the coupling of the atoms, the ground state density $n_{\sigma}(\bm{r})$ is formed as shown on the left of the figure. It is clearly seen that the electron population at the atomic core is increased by the formation of the covalent bond although the overall distribution is not significantly changed.    
On the other hand, when the bond is stretched to $R$(H-H)$\;=7.0$ a.u., the ground state density $n_{\sigma}(\bm{r})$ becomes totally different from the reference density as shown in Fig. \ref{fig:density_7au}. Actually, this plot is nothing more than a realistic representation of Fig. \ref{fig:TF_model}. The density of the entangled state is characterized by a completely delocalized distribution over two distant atomic sites, which causes the error in the description of the bond dissociation using the approximate density functional based on the homogeneous electron gas.       

\begin{figure}[h]
\centering
\scalebox{0.32}[0.32] {\includegraphics[trim=20 170 0 120,clip]{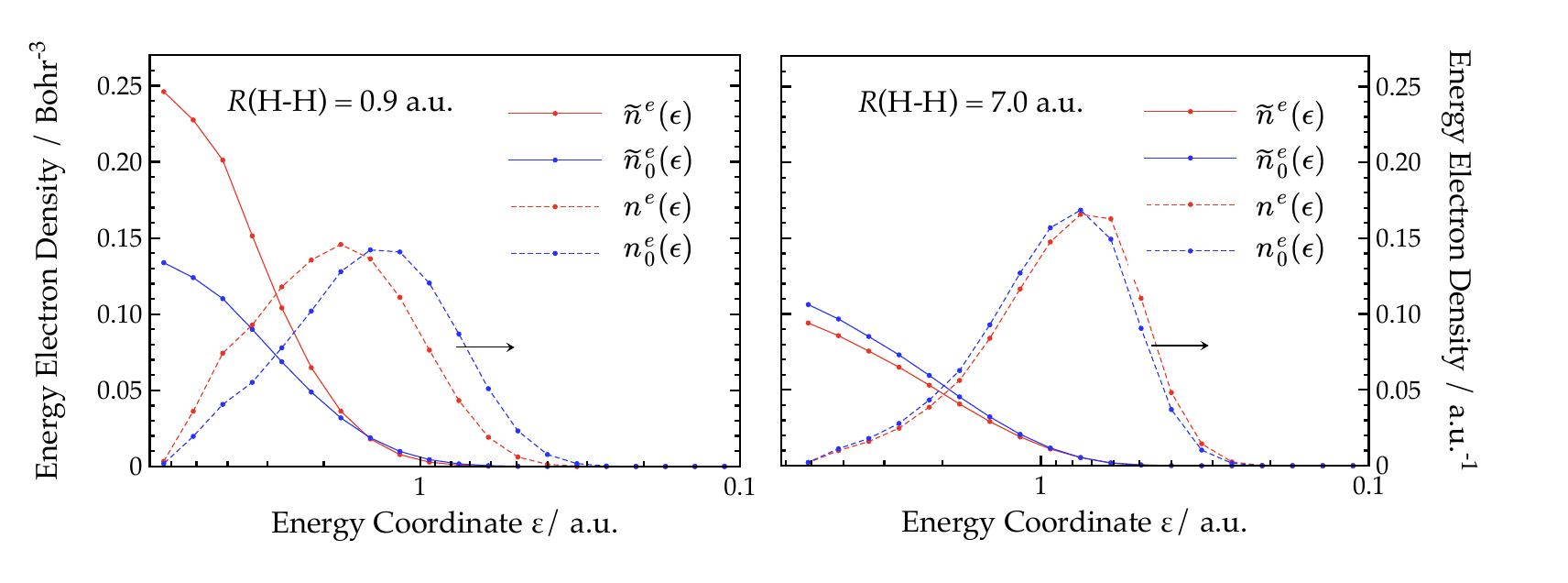}}      
\caption{\label{fig:ene_dnst} Energy electron densities $\widetilde{n}^e(\epsilon)$ and $n^e(\epsilon)$ for each spin in the H$_2$ molecule at $R$(H-H)$\;=0.9$ (Left) and $7.0$ a.u. (Right). The subscript $0$ indicates the reference density. The broken lines refer to the axis on the right. }
\end{figure}

It is useful to examine what happens when the electron densities $n(\bm{r})$ and $n_0(\bm{r})$ are projected onto the energy coordinate $\epsilon$ to yield the energy electron distributions.  In Fig. \ref{fig:ene_dnst}, the energy electron densities $n^e(\epsilon)$ and $\widetilde{n}^e(\epsilon)$ defined, respectively, in Eqs. (\ref{eq:dnst_e}) and (\ref{eq:tilde_dnst_e}) are presented for $R(\rm{H-H}) = 0.9$ and $7.0$ a.u. We note that the dimension of $\widetilde{n}^e(\epsilon)$ is volume$^{-1}$ and is the same as that of the normal electron density, while $n^e(\epsilon)$ has the dimension of energy$^{-1}$ by definitions. Since the volume $\Omega(\epsilon)$ at the largest energy coordinate (the left end of the horizontal axis) is close to zero, $n^e(\epsilon)$ becomes almost $0$ at the left end. At $R(\rm{H-H}) = 0.9$ a.u., which is much shorter than the equilibrium bond distance $R_{\rm{eq}} = 1.4$ a.u. of H$_2$, the energy electron density $\widetilde{n}^e(\epsilon)$ of the system with the ground state density $n(\bm{r})$ is remarkably different from that of the reference density $n_0(\bm{r})$. A comparison of $\widetilde{n}^e(\epsilon)$ with $\widetilde{n}_0^e(\epsilon)$ suggests that the electron distribution at the lower energy coordinate moves toward the regions with the higher energy coordinate associated with the variation of $n_0 \to n$. This trend can also be confirmed by comparing $\widetilde{n}^e(\epsilon)$ with $\widetilde{n}_0^e(\epsilon)$. In contrast, at $R(\rm{H-H}) = 7.0$ a.u., where the covalent bond is completely broken, the energy electron density for the ground state density $n(\bm{r})$ is almost coincident with that for the reference density. 
In fact, this property of the energy distributions is favorable for the development of the approximate density functional $E^e[n^e]$, since the relation of $E^{e}\left[n^{e}\right]=E^{e}\left[n_{0}^{e}\right]$ is guaranteed at the dissociation limit under any choice of the functional $E^e$. On the contrary, it seems to be very difficult to develop the normal density functional $E[n]$ which allows to correctly describe the bond dissociation because the functional has to realize the relation $E\left[n\right]=E\left[n_{0}\right]$, although $n(\bm{r})$ is completely different from $n_0(\bm{r})$ as shown in Fig. \ref{fig:density_7au}. 

\subsection{Potential Energy Curve of Hydrogen Molecule}    
\begin{figure}[h]
\centering
\scalebox{0.3}[0.3] {\includegraphics[trim=0 0 0 0,clip]{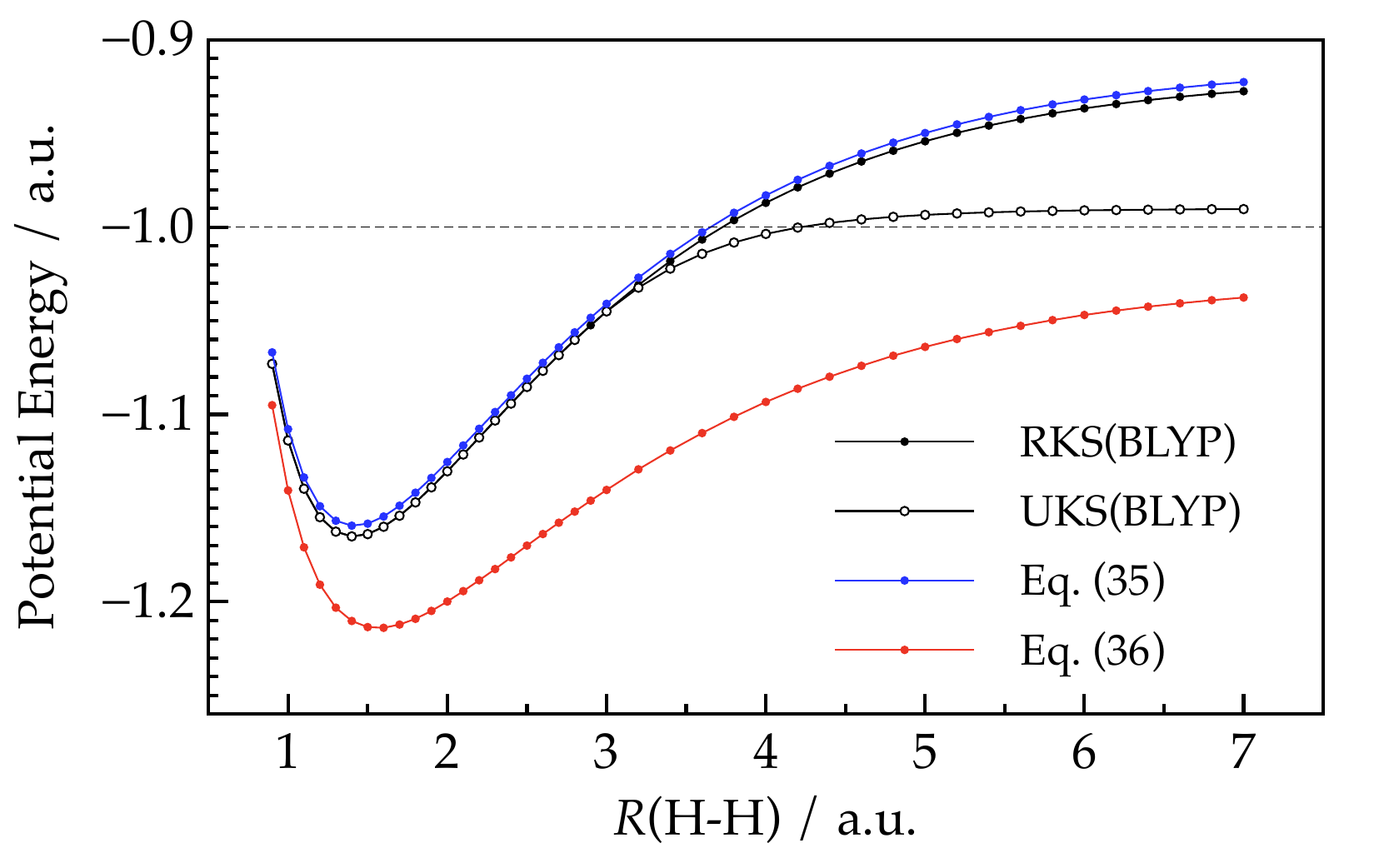}}      
\caption{\label{fig:H2_Ex_SC} Potential energy curves of H$_2$ molecule computed using Eqs. (\ref{eq:EKS_without_sc}) and (\ref{eq:EKS_with_sc}). The results of the spin-restricted (RKS) and spin-unrestricted (UKS) Kohn-Sham DFT calculations with the BLYP functional are also depicted. The horizontal broken line shows the correct asymptotic energy at the dissociation. }
\end{figure}

The potential energy curves (PECs) of the H$_2$ molecule are evaluated using Eq. (\ref{eq:E_total}), which includes the LDA exchange energy $\overline{E}_x^{\rm{LDA}}[n]$ with the static correlation as shown in Eq. (\ref{eq:Ex_sc}). Note that the kinetic energy functional Eq. (\ref{eq:Ekin_func}) also includes the static correlation $E_{\rm{kin}}^{\rm{sc}}[n_0]$ defined in Eq. (\ref{eq:Ekin_sc}). In order to discuss separately the effect of the $E_{\rm{kin}}^{\rm{sc}}[n_0]$ from the static correlation in the exchange correlation energy, we first present the various PECs computed by the KS-DFT with and without the static correlation energy in the term $\overline{E}_x^{\rm{LDA}}[n]$ defined in Eq. (\ref{eq:Ex_sc}). Explicitly, we examined the following two energy functionals 
\begin{eqnarray}
E_{{\rm tot}}^{{\rm KS}}\left[\left\{ \varphi_{i}\right\} \right] & =E_{{\rm kin}}^{{\rm KS}}\left[\left\{ \varphi_{i}\right\} \right]+E_{x}^{e,{\rm LDA}}\left[n^e\right]       \nonumber    \\
 & +E_{{\rm BLYP}}^{\prime}\left[n\right]+E_{{\rm H}}\left[n\right]+E_{{\rm ext}}\left[n\right]
\label{eq:EKS_without_sc}
\end{eqnarray}
and
\begin{eqnarray}
E_{{\rm tot}}^{{\rm KS,SC}}\left[\left\{ \varphi_{i}\right\} \right] & =E_{{\rm kin}}^{{\rm KS}}\left[\left\{ \varphi_{i}\right\} \right]+\overline{E}_{x}^{{\rm LDA}}\left[n\right]     \nonumber \\
 & +E_{{\rm BLYP}}^{\prime}\left[n\right]+E_{{\rm H}}\left[n\right]+E_{{\rm ext}}\left[n\right]
\label{eq:EKS_with_sc} 
\end{eqnarray}
where $\{\varphi_i\}$ are the KS wave functions determined by Eq. (\ref{eq:KSeq}), and $E_{{\rm kin}}^{{\rm KS}}\left[\left\{ \varphi_{i}\right\} \right]$ is computed using the KS wave functions. Note that the difference between Eqs. (\ref{eq:EKS_with_sc}) and (\ref{eq:EKS_without_sc}) is the exchange static correlation $\overline{E}_{x}^{{\rm LDA}}\left[n\right]-E_{x}^{e,{\rm LDA}}\left[n^e\right]$ = $E_x^{\rm LDA}[n_0]-E_x^{e,{\rm LDA}}[n_0^e]$ from the definition of Eq. (\ref{eq:Ex_sc}). It is also important to note that Eq. (\ref{eq:EKS_without_sc}) becomes identical to the conventional DFT energy with the BLYP functional when $E_{x}^{e,{\rm LDA}}\left[n^e\right]$ is replaced by the normal LDA energy $E_{x}^{\rm LDA}\left[n\right]$.  The results are presented in Fig. \ref{fig:H2_Ex_SC}. To make comparisons, the spin-restricted (RKS) and spin-unrestricted (UKS) KS-DFT calculations with the BLYP functional are also performed. In the graph, the UKS calculation gives the dissociation energy $D_e = 109.7$ kcal/mol, showing a good agreement with an experimental value(109 kcal/mol).    
At first sight, it is quite impressive that the PEC given by Eq. (\ref{eq:EKS_without_sc}) shows a nice agreement with that by the RKS calculation. It directly implies that projecting the density $n(\bm{r})$ onto the energy coordinate $\epsilon$ using Eq. (\ref{eq:dnst_e}) hardly degrades the LDA energy. A similar result has already been reported in Ref. \cite{rf:Takahashi2018} for a more complicated exchange functional based on the Becke-Roussel approach\cite{rf:becke1989pra}. In any case, Eq. (\ref{eq:EKS_without_sc}) as well as the RKS calculation overshoot the energy at the bond dissociation ($R$(H-H) $ = 7.0$\; a.u.) because these methods do not include the static correlation functional. In contrast, the PEC by Eq. (\ref{eq:EKS_with_sc}) provides the lower energy than the RKS due to the static correlation $E_x^{\rm LDA}[n_0]-E_x^{e,{\rm LDA}}[n_0^e]$. However, the static correlation evaluated in Eq. (\ref{eq:EKS_with_sc}) seems to be too strong especially in the middle range of the bond distance. It should also be noted that the underestimation of the energy by Eq. (\ref{eq:EKS_with_sc}) at the dissociation limit is partly due to the LYP correlation energy. The LYP functional is designed so that the correlation energy completely vanishes for an electron with  spin $\sigma$. Thus, at the dissociation limit, the UKS provides the correct correlation energy. On the other hand, in the RKS calculation, the LYP energy remains even at the dissociation limit. Actually, the LYP energy is found to be $-0.027$ a.u. at $R$(H-H)$\;=7.0$ a.u. In Ref. \cite{Takahashi2020jpb}, HT proposed a method to attenuate the static correlation in the middle region of the bond distance according to the degree of the delocalization of the exchange hole. In the present work, however, the method is not applied because it is expected that substantial error cancellation will take place between the static correlations in the exchange and the kinetic energies. 

\begin{figure}[h]
\centering
\scalebox{0.3}[0.3] {\includegraphics[trim=0 0 0 0,clip]{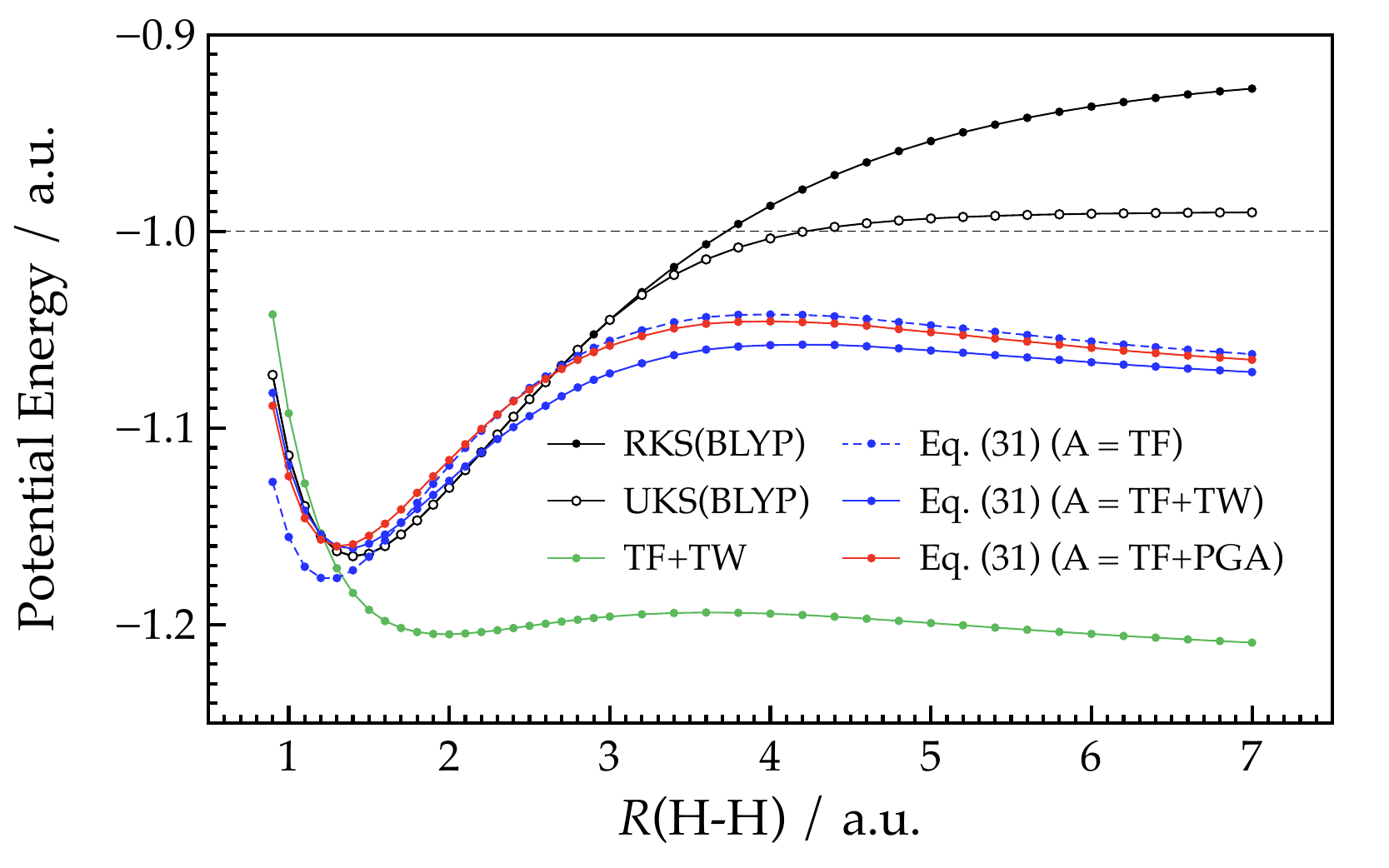}}      
\caption{\label{fig:H2_dissoc} Potential energy curves of H$_2$ molecule computed using Eq. (\ref{eq:E_total}). The notations \textquoteleft TF\textquoteright , \textquoteleft TF+TW\textquoteright\; and \textquoteleft TF+PGA\textquoteright\; in the parentheses indicate the methods specified by Eqs. (\ref{eq:SC_TF}), (\ref{eq:SC_TF_TW}) and (\ref{eq:SC_TF_PGA}), respectively. \textquoteleft TF+TW\textquoteright\; without a parenthesis indicates the normal TF energy with a GGA correction by Tran and Weso\l owski (TW)\cite{Tran2002ijqc}.  The results of the spin-restricted (RKS) and spin-unrestricted (UKS) Kohn-Sham DFT calculations with the BLYP functional are also depicted. In the calculations of the kinetic energy functionals, the electron density obtained by the RKS calculation is used. The horizontal broken line shows the correct asymptotic energy at the dissociation. }
\end{figure} 
In Fig. \ref{fig:H2_dissoc}, we present the PECs of the H$_2$ molecule obtained using Eq. (\ref{eq:E_total}). For comparisons, the conventional TF model with the GGA correction by Tran and Weso\l owski (TW)\cite{Tran2002ijqc} is also used to evaluate the kinetic energy (shown with the legend \textquoteleft TF+TW\textquoteright\;).  The result given by  \textquoteleft TF+TW\textquoteright\; shows a rather good behavior only in the region of the short bond distance ($R$(H-H)$\;\leq 1.3$\; a.u.). In the region of longer bond distances, however, it is found that the PEC seriously underestimates the kinetic energy. The mechanism underlying the failure has already been discussed in subsection 2.1. By the calculation with the TF model (not shown in the graph),  we found that the PEC is shifted by $\sim0.14$ a.u. downward at $R$(H-H)$\;\leq 0.9$\; a.u. due to the serious underestimation of the kinetic energy. One might consider that the kinetic energy \textquoteleft TF+TW\textquoteright\; will be raised for the longer bond distances when the spin-polarized density is used as the argument. However, since the spin-polarization takes place around $R$(H-H) = 3.0 a.u., the kinetic energies in the region from $R$(H-H) = 1.4 to 3.0 a.u. will not be improved even when the spin polarization is allowed.  

It is quite striking that the inclusion of the kinetic static correlation $E_{\rm{kin}}^{\rm{sc}}[n_0]$ defined in Eq. (\ref{eq:Ekin_sc}) offers a drastic improvement of the PEC as shown in the graph with the legend  \textquoteleft Eq. (\ref{eq:E_total}) (A = TF)\textquoteright . Actually, it is observed that a well-structured potential well is formed. However, it is shown that the energy in the region of longer bond distances is still underestimated. The energy at $R$(H-H)$\;\leq 0.9$\; a.u. is also underestimated by $0.05$ a.u., which leads to a shortening of the equilibrium bond distance $R_{\rm{eq}}$ from $1.4$ to $1.3$ a.u. 
It is demonstrated in the graph \textquoteleft Eq. (\ref{eq:E_total}) (A = TF+TW)\textquoteright\; that the application of the GGA correction by TW to the coupling parameter integration in Eq. (\ref{eq:coupling_int}) favorably raises the energies of \textquoteleft A = TF\textquoteright\; in the region of the bond distance $R$(H-H)$\;\leq 1.6$\; a.u. As a result, the PEC of \textquoteleft A = TF+TW\textquoteright\; around $R_{\rm{eq}}$ becomes almost coincident with that of the RKS calculation. It is found that this functional gives the dissociation energy of $56.3$ kcal/mol at $R$(H-H) = 7.0 a.u. although the energy is seemingly not converged.    

We also test the potential gradient approximation (PGA) described in subsection 2.3 instead of the GGA correction. The corresponding graph is indicated by the legend \textquoteleft Eq. (\ref{eq:E_total}) (A = TF+PGA)\textquoteright\; in Fig. \ref{fig:H2_dissoc}. Since this might be the first application of the PGA to the TF model, it is helpful to consider how the approximation improves the TF energy describing the bond dissociation. When the bond distance is short, the Coulomb  potential of a hydrogen atom overlaps significantly with that of the other hydrogen atom, which causes the increase in the potential gradient $q(\bm{r})\equiv |\nabla \upsilon_{\rm{ext}}(\bm{r})|$ in Eq. (\ref{eq:PG_e}). Thus, the PGA correction to the TF energy becomes larger as $R$(H-H) becomes shorter. Consequently, the graph \textquoteleft A = TF+PGA\textquoteright\; shows a good agreement with the RKS potential in the region of $R$(H-H)$\;\leq 1.3$\; a.u. In the present application, the fitting parameters $(\beta, \gamma)$ in Eq. (\ref{eq:PGA_enhcf_e}) have been roughly chosen as $\beta = \gamma = 4.0\times10^{-3}$ since the parameter optimization is not the major purpose of the work. It is expected that these parameters will need to be modified slightly depending on the system of interest. In any case, it is demonstrated that the PGA approach shows a performance comparable to a GGA functional, at least for the H$_2$ molecule. 

\begin{figure}[h]
\centering
\scalebox{0.3}[0.3] {\includegraphics[trim=0 0 0 0,clip]{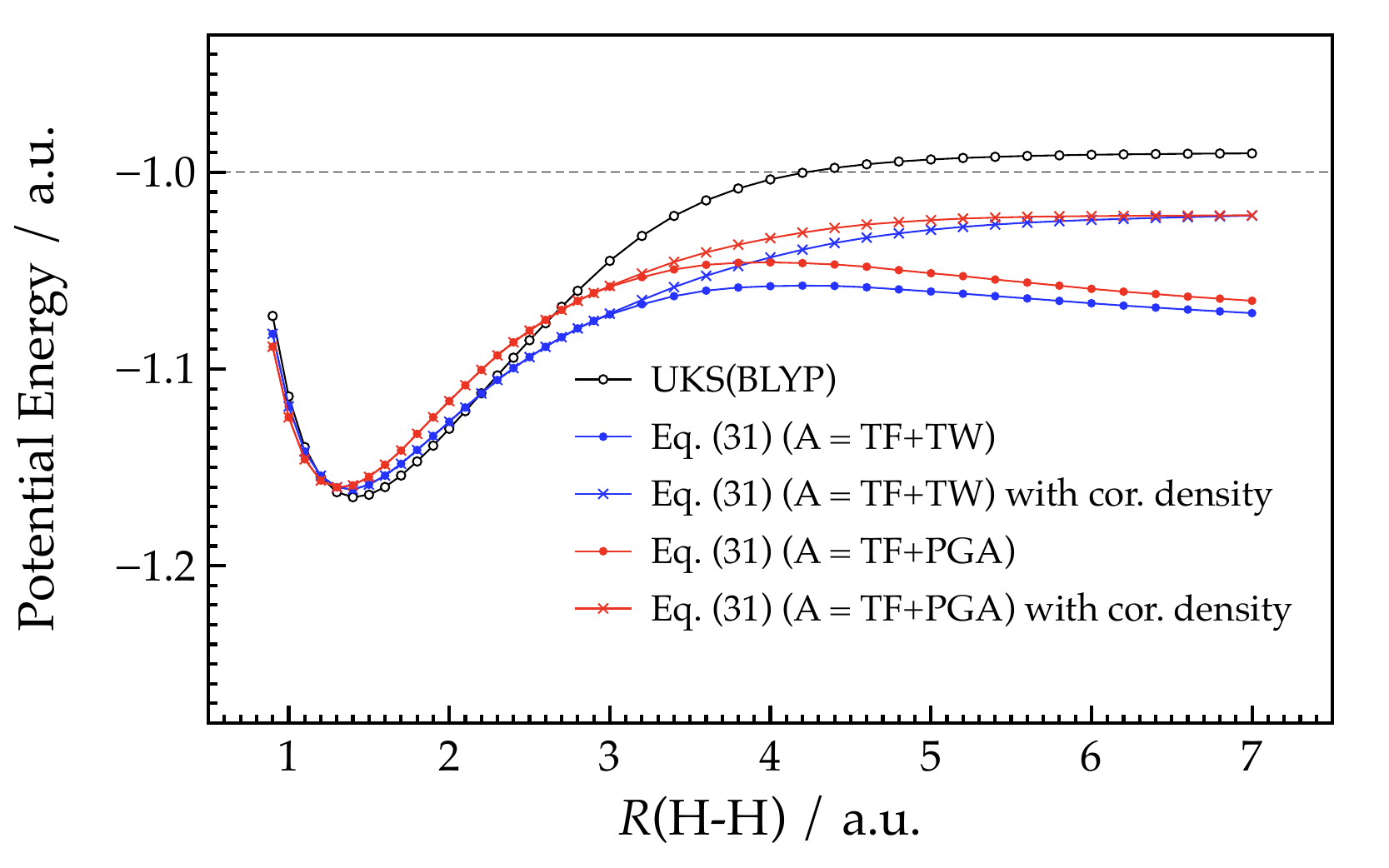}}      
\caption{\label{fig:H2_dissoc_2} Potential energy curves of H$_2$ molecule computed using Eq. (\ref{eq:E_total}). The notations \textquoteleft TF+TW\textquoteright\; and \textquoteleft TF+PGA\textquoteright\; in the parentheses indicate the methods specified by Eqs. (\ref{eq:SC_TF_TW}) and (\ref{eq:SC_TF_PGA}), respectively. The definition of the corrected density is provided in Eq. (\ref{eq:corrected_dnst}). The result of the spin-unrestricted (UKS) Kohn-Sham DFT calculation with the BLYP functional is also depicted.  The horizontal broken line shows the correct asymptotic energy at the dissociation. }
\end{figure}

Finally, we close this section with a discussion of the lowering of the energies by the kinetic energy functional (KEF) of Eq. (\ref{eq:E_total}) in the region of the long bond distances. Actually, at $R$(H-H)$\;= 7.0$ a.u. the total energy is underestimated by about $-0.07$ a.u. by the KEF of Eq. (\ref{eq:E_total}). As described above, the part of the error comes from the LYP correlation energy evaluated as $-0.027$ a.u. for the spin-adapted electron density. We attribute the source of the remaining error to the broadening of the RKS density ($n^{\rm{RKS}} = n_{\alpha}^{\rm{RKS}} = n_{\beta}^{\rm{RKS}}$) at each site upon the bond dissociation as compared to the \textit{total} UKS density. To examine this effect, we introduce the density which gradually varies from the density $n^{{\rm RKS}}$ of the RKS wave function to that of the UKS. Explicitly, the density for each spin is written as
\begin{equation}
\overline{n}\left(\bm{r}\right)=\left(1-c\right)n^{{\rm RKS}}\left(\bm{r}\right)+c\;\overline{n}^{{\rm UKS}}\left(\bm{r}\right)
\label{eq:corrected_dnst}
\end{equation}    
with $c$ being a real coupling constant satisfying $0\leq c\leq1$.  The spin-adapted UKS density $\overline{n}^{{\rm UKS}}$ is 
given by
\begin{equation}
\overline{n}^{{\rm UKS}}\left(\bm{r}\right)=\frac{1}{2}\left(n_{\alpha}^{{\rm UKS}}\left(\bm{r}\right)+n_{\beta}^{{\rm UKS}}\left(\bm{r}\right)\right)
\end{equation}    
It is assumed that the coupling constant $c$ is linearly dependent on the distance $R\equiv R$(H-H), thus,
\begin{equation}
c=\frac{R-R_{{\rm min}}}{R_{{\rm max}}-R_{{\rm min}}}\quad(R_{{\rm min}}\leq R\leq R_{{\rm max}})
\end{equation}
with $R_{{\rm min}} = 2.0$ a.u. and $R_{{\rm max}} = 7.0$ a.u. For the region of $R < 2.0$ a.u., $c$ is assumed to be 0. Then, the \textit{corrected} density is substituted to the functionals of Eq. (\ref{eq:E_total}). The results are presented in Fig. \ref{fig:H2_dissoc_2}. As clearly shown in the figure, the decrease in the energies at the dissociation limit is suppressed by the introduction of the corrected density. It is thus proved that the use of the RKS density leads to the stabilization of the kinetic energy due to its delocalized nature as compared to the UKS density. Anyway, the dissociation energy with the corrected density is evaluated as $87.4$ kcal/mol.   
 
\section{Conclusion}
In this work, a kinetic energy density functional (KEF) was developed to describe the dissociation of a covalent bond in a molecule. The difficulty of the Thomas-Fermi (TF) model or the TF with a GGA correction in calculating the bond dissociation arises from the fact that the kinetic energy density of the TF model is seriously decreased due to the delocalization of the electron density upon dissociation. Therefore, from the numerical point of view, the source of the error associated with these method is essentially the same as that of the static correlation error in the local density approximation (LDA). Based on this consideration, the KEF is formulated with the coupling-parameter integration scheme in parallel to the formulation of the exchange correlation functional\cite{rf:Takahashi2018, Takahashi2020jpb} that includes the static correlation. The essential feature of the functional is that it uses the energy electron density $n^e(\epsilon)$ as a fundamental variable on the basis of the rigorous framework of the novel DFT\cite{rf:Takahashi2018}. The new KEF exploits an interesting property specific to the distribution $n^e(\epsilon)$, that is, $n_\sigma^e(\epsilon)$ for the dissociated entangled system becomes identical to that for the isolated atom in the system. We also developed a potential gradient approximation (PGA) instead of the GGA to improve the TF model.       

A series of the new KEFs was applied to the dissociation of an H$_2$ molecule. The exchange-correlation functional with the static correlation developed in Ref. \cite{rf:Takahashi2018} is also used in combination with the KEF. It was demonstrated that the overall behavior of the potential energy curve (PEC) of H$_2$ can be correctly described with the KEFs. It was also clearly shown that the GGA correction to the functional $E_{\rm{kin}}^{e,\rm{TF}}[n^e]$ favorably improves the PEC of the molecule near the equilibrium bond distance. The PGA approach was also applied to the functional $E_{\rm{kin}}^{e,\rm{TF}}[n^e]$ and it was shown that the approximation offers an improvement comparable to the GGA functional.    

In the new DFT, where the energy electron density $n_\sigma^e(\epsilon)$ plays a fundamental role, the electrons that feel the same Coulomb energy from the nuclei are considered as a single entity as a whole, whereby the problem arising from the severe delocalization of the spatial electron density $n(\bm{r})$ can be solved in a natural way. 
It is thus expected that the new DFT using the density $n_\sigma^e(\epsilon)$ offers a potential route to the development of the KEF applicable to chemical events.  The KEF developed in this work should also be applied to the dissociation of multiple bonds in the near future. More importantly, the functional must be extended so that it can be used in the self-consistent field calculation for the density. We anticipate that some nonlocal term built from a response function\cite{Takahashi2022ijqc} must be included in the functional. These works are now undertaken and will be reported elsewhere.       \\  

\begin{acknowledgments}
This paper was supported by the Grant-in-Aid for Scientific Research(C) (Nos. 22K12055, 17K05138) from the Japan Society for the Promotion of Science (JSPS); the Grant-in-Aid for Scientific Research on Innovative Areas (No. 23118701) from the Ministry of Education, Culture, Sports, Science, and Technology (MEXT); the Grant-in-Aid for Challenging Exploratory Research (No. 25620004) from the Japan Society for the Promotion of Science (JSPS). This research also used computational resources of the HPCI system provided by Kyoto, Nagoya, and Osaka university through the HPCI System Research Project (Project IDs: hp170046, hp180030, hp180032, hp190011, and hp200016). 
\end{acknowledgments}  


\begin{thebibliography}{42}%
\makeatletter
\providecommand \@ifxundefined [1]{%
 \@ifx{#1\undefined}
}%
\providecommand \@ifnum [1]{%
 \ifnum #1\expandafter \@firstoftwo
 \else \expandafter \@secondoftwo
 \fi
}%
\providecommand \@ifx [1]{%
 \ifx #1\expandafter \@firstoftwo
 \else \expandafter \@secondoftwo
 \fi
}%
\providecommand \natexlab [1]{#1}%
\providecommand \enquote  [1]{``#1''}%
\providecommand \bibnamefont  [1]{#1}%
\providecommand \bibfnamefont [1]{#1}%
\providecommand \citenamefont [1]{#1}%
\providecommand \href@noop [0]{\@secondoftwo}%
\providecommand \href [0]{\begingroup \@sanitize@url \@href}%
\providecommand \@href[1]{\@@startlink{#1}\@@href}%
\providecommand \@@href[1]{\endgroup#1\@@endlink}%
\providecommand \@sanitize@url [0]{\catcode `\\12\catcode `\$12\catcode
  `\&12\catcode `\#12\catcode `\^12\catcode `\_12\catcode `\%12\relax}%
\providecommand \@@startlink[1]{}%
\providecommand \@@endlink[0]{}%
\providecommand \url  [0]{\begingroup\@sanitize@url \@url }%
\providecommand \@url [1]{\endgroup\@href {#1}{\urlprefix }}%
\providecommand \urlprefix  [0]{URL }%
\providecommand \Eprint [0]{\href }%
\providecommand \doibase [0]{https://doi.org/}%
\providecommand \selectlanguage [0]{\@gobble}%
\providecommand \bibinfo  [0]{\@secondoftwo}%
\providecommand \bibfield  [0]{\@secondoftwo}%
\providecommand \translation [1]{[#1]}%
\providecommand \BibitemOpen [0]{}%
\providecommand \bibitemStop [0]{}%
\providecommand \bibitemNoStop [0]{.\EOS\space}%
\providecommand \EOS [0]{\spacefactor3000\relax}%
\providecommand \BibitemShut  [1]{\csname bibitem#1\endcsname}%
\let\auto@bib@innerbib\@empty
\bibitem [{\citenamefont {Parr}\ and\ \citenamefont
  {Yang}(1989)}]{rf:parr_yang_eng}%
  \BibitemOpen
  \bibfield  {author} {\bibinfo {author} {\bibfnamefont {R.~G.}\ \bibnamefont
  {Parr}}\ and\ \bibinfo {author} {\bibfnamefont {W.}~\bibnamefont {Yang}},\
  }\href@noop {} {\emph {\bibinfo {title} {Density-functional theory of atoms
  and molecules}}}\ (\bibinfo  {publisher} {Oxford university press},\ \bibinfo
  {address} {New York},\ \bibinfo {year} {1989})\BibitemShut {NoStop}%
\bibitem [{\citenamefont {Hohenberg}\ and\ \citenamefont
  {Kohn}(1964)}]{hohenberg1964}%
  \BibitemOpen
  \bibfield  {author} {\bibinfo {author} {\bibfnamefont {P.}~\bibnamefont
  {Hohenberg}}\ and\ \bibinfo {author} {\bibfnamefont {W.}~\bibnamefont
  {Kohn}},\ }\bibfield  {title} {\bibinfo {title} {Inhomogeneous electron
  gas},\ }\href@noop {} {\bibfield  {journal} {\bibinfo  {journal} {Phys.
  Rev.}\ }\textbf {\bibinfo {volume} {136}},\ \bibinfo {pages} {B864} (\bibinfo
  {year} {1964})}\BibitemShut {NoStop}%
\bibitem [{\citenamefont {Kohn}\ and\ \citenamefont
  {Sham}(1965)}]{rf:kohn1965pr}%
  \BibitemOpen
  \bibfield  {author} {\bibinfo {author} {\bibfnamefont {W.}~\bibnamefont
  {Kohn}}\ and\ \bibinfo {author} {\bibfnamefont {L.~J.}\ \bibnamefont
  {Sham}},\ }\bibfield  {title} {\bibinfo {title} {{Self-consistent equations
  including exchange and correlation effects}},\ }\href@noop {} {\bibfield
  {journal} {\bibinfo  {journal} {Phys. Rev.}\ }\textbf {\bibinfo {volume}
  {140}},\ \bibinfo {pages} {A1133} (\bibinfo {year} {1965})}\BibitemShut
  {NoStop}%
\bibitem [{\citenamefont {Wesolowski}\ and\ \citenamefont
  {Wang}(2013)}]{Wesolowski2013}%
  \BibitemOpen
  \bibinfo {editor} {\bibfnamefont {T.~A.}\ \bibnamefont {Wesolowski}}\ and\
  \bibinfo {editor} {\bibfnamefont {Y.~A.}\ \bibnamefont {Wang}},\ eds.,\
  \href@noop {} {\emph {\bibinfo {title} {Recent Progress in Orbital Free
  Density Functional Theory (Recent Advances in Computational Chemistry)}}}\
  (\bibinfo  {publisher} {World Scientific},\ \bibinfo {address} {Singapore},\
  \bibinfo {year} {2013})\BibitemShut {NoStop}%
\bibitem [{\citenamefont {Hung}\ \emph {et~al.}(2010)\citenamefont {Hung},
  \citenamefont {Huang}, \citenamefont {Shin}, \citenamefont {Ho},
  \citenamefont {Ligneres},\ and\ \citenamefont {Carter}}]{rf:hung2010cpc}%
  \BibitemOpen
  \bibfield  {author} {\bibinfo {author} {\bibfnamefont {L.}~\bibnamefont
  {Hung}}, \bibinfo {author} {\bibfnamefont {C.}~\bibnamefont {Huang}},
  \bibinfo {author} {\bibfnamefont {I.}~\bibnamefont {Shin}}, \bibinfo {author}
  {\bibfnamefont {G.~S.}\ \bibnamefont {Ho}}, \bibinfo {author} {\bibfnamefont
  {V.~L.}\ \bibnamefont {Ligneres}},\ and\ \bibinfo {author} {\bibfnamefont
  {E.~A.}\ \bibnamefont {Carter}},\ }\bibfield  {title} {\bibinfo {title}
  {Introducing profess 2.0: a parallelized, fully linear scaling program for
  orbital-free density functional theory calculations},\ }\href@noop {}
  {\bibfield  {journal} {\bibinfo  {journal} {Comput. Phys. Commun.}\ }\textbf
  {\bibinfo {volume} {181}},\ \bibinfo {pages} {2208} (\bibinfo {year}
  {2010})}\BibitemShut {NoStop}%
\bibitem [{\citenamefont {Thomas}(1927)}]{Thomas1927PCPS}%
  \BibitemOpen
  \bibfield  {author} {\bibinfo {author} {\bibfnamefont {L.~H.}\ \bibnamefont
  {Thomas}},\ }\bibfield  {title} {\bibinfo {title} {The calculation of atomic
  fields},\ }\href@noop {} {\bibfield  {journal} {\bibinfo  {journal} {Proc.
  Camb. Phil. Soc.}\ }\textbf {\bibinfo {volume} {23}},\ \bibinfo {pages} {541}
  (\bibinfo {year} {1927})}\BibitemShut {NoStop}%
\bibitem [{\citenamefont {Fermi}(1928)}]{Fermi1928zp}%
  \BibitemOpen
  \bibfield  {author} {\bibinfo {author} {\bibfnamefont {E.}~\bibnamefont
  {Fermi}},\ }\bibfield  {title} {\bibinfo {title} {A statistical method for
  the determination of some atomic properties and the application of this
  method to the theory of the periodic system of elements},\ }\href@noop {}
  {\bibfield  {journal} {\bibinfo  {journal} {Z. Phys.}\ }\textbf {\bibinfo
  {volume} {48}},\ \bibinfo {pages} {73} (\bibinfo {year} {1928})}\BibitemShut
  {NoStop}%
\bibitem [{\citenamefont {Bal\`{a}zs}(1967)}]{Balazs1967pr}%
  \BibitemOpen
  \bibfield  {author} {\bibinfo {author} {\bibfnamefont {N.~L.}\ \bibnamefont
  {Bal\`{a}zs}},\ }\bibfield  {title} {\bibinfo {title} {Formation of stable
  molecules within the statistical theory of atoms},\ }\href@noop {} {\bibfield
   {journal} {\bibinfo  {journal} {Phys. Rev.}\ }\textbf {\bibinfo {volume}
  {156}},\ \bibinfo {pages} {42} (\bibinfo {year} {1967})}\BibitemShut
  {NoStop}%
\bibitem [{\citenamefont {Hodges}(1973)}]{Hodges1973CanJChem}%
  \BibitemOpen
  \bibfield  {author} {\bibinfo {author} {\bibfnamefont {C.~H.}\ \bibnamefont
  {Hodges}},\ }\bibfield  {title} {\bibinfo {title} {Quantum corrections to the
  thomas--fermi approximation---the kirzhnits method},\ }\href@noop {}
  {\bibfield  {journal} {\bibinfo  {journal} {Can. J. Chem}\ }\textbf {\bibinfo
  {volume} {51}},\ \bibinfo {pages} {1428} (\bibinfo {year}
  {1973})}\BibitemShut {NoStop}%
\bibitem [{\citenamefont {Perdew}\ \emph {et~al.}(1988)\citenamefont {Perdew},
  \citenamefont {Levy}, \citenamefont {Painter}, \citenamefont {Wei},\ and\
  \citenamefont {Lagowski}}]{Perdew1988prb}%
  \BibitemOpen
  \bibfield  {author} {\bibinfo {author} {\bibfnamefont {J.~P.}\ \bibnamefont
  {Perdew}}, \bibinfo {author} {\bibfnamefont {M.}~\bibnamefont {Levy}},
  \bibinfo {author} {\bibfnamefont {G.~S.}\ \bibnamefont {Painter}}, \bibinfo
  {author} {\bibfnamefont {S.}~\bibnamefont {Wei}},\ and\ \bibinfo {author}
  {\bibfnamefont {J.~B.}\ \bibnamefont {Lagowski}},\ }\bibfield  {title}
  {\bibinfo {title} {Chemical bond as a test of density-gradient expansions for
  kinetic and exchange energies},\ }\href@noop {} {\bibfield  {journal}
  {\bibinfo  {journal} {Phys. Rev. B}\ }\textbf {\bibinfo {volume} {37}},\
  \bibinfo {pages} {838} (\bibinfo {year} {1988})}\BibitemShut {NoStop}%
\bibitem [{\citenamefont {Lee}\ \emph {et~al.}(1991)\citenamefont {Lee},
  \citenamefont {Lee},\ and\ \citenamefont {Parr}}]{Lee1991pra}%
  \BibitemOpen
  \bibfield  {author} {\bibinfo {author} {\bibfnamefont {H.}~\bibnamefont
  {Lee}}, \bibinfo {author} {\bibfnamefont {C.}~\bibnamefont {Lee}},\ and\
  \bibinfo {author} {\bibfnamefont {R.~G.}\ \bibnamefont {Parr}},\ }\bibfield
  {title} {\bibinfo {title} {Conjoint gradient correction to the hartree-fock
  kinetic- and exchange-energy density functionals},\ }\href@noop {} {\bibfield
   {journal} {\bibinfo  {journal} {Phys. Rev. A}\ }\textbf {\bibinfo {volume}
  {44}},\ \bibinfo {pages} {768} (\bibinfo {year} {1991})}\BibitemShut
  {NoStop}%
\bibitem [{\citenamefont {Tran}\ and\ \citenamefont
  {Weso{\l}owski}(2002)}]{Tran2002ijqc}%
  \BibitemOpen
  \bibfield  {author} {\bibinfo {author} {\bibfnamefont {F.}~\bibnamefont
  {Tran}}\ and\ \bibinfo {author} {\bibfnamefont {T.~A.}\ \bibnamefont
  {Weso{\l}owski}},\ }\bibfield  {title} {\bibinfo {title} {Link between the
  kinetic- and exchange-energy functionals in the generalized gradient
  approximation},\ }\href@noop {} {\bibfield  {journal} {\bibinfo  {journal}
  {Int. J. Quantum Chem.}\ }\textbf {\bibinfo {volume} {89}},\ \bibinfo {pages}
  {441} (\bibinfo {year} {2002})}\BibitemShut {NoStop}%
\bibitem [{\citenamefont {Thakkar}(1992)}]{Thakkar1992}%
  \BibitemOpen
  \bibfield  {author} {\bibinfo {author} {\bibfnamefont {A.~J.}\ \bibnamefont
  {Thakkar}},\ }\bibfield  {title} {\bibinfo {title} {{Comparison of
  kinetic-energy density functionals}},\ }\href
  {https://doi.org/10.1103/PhysRevA.46.6920} {\bibfield  {journal} {\bibinfo
  {journal} {Phys. Rev. A}\ }\textbf {\bibinfo {volume} {46}},\ \bibinfo
  {pages} {6920} (\bibinfo {year} {1992})}\BibitemShut {NoStop}%
\bibitem [{\citenamefont {Vitos}\ \emph {et~al.}(2000)\citenamefont {Vitos},
  \citenamefont {Johansson}, \citenamefont {Kollar},\ and\ \citenamefont
  {Skriver}}]{rf:vitos2000pra}%
  \BibitemOpen
  \bibfield  {author} {\bibinfo {author} {\bibfnamefont {L.}~\bibnamefont
  {Vitos}}, \bibinfo {author} {\bibfnamefont {B.}~\bibnamefont {Johansson}},
  \bibinfo {author} {\bibfnamefont {J.}~\bibnamefont {Kollar}},\ and\ \bibinfo
  {author} {\bibfnamefont {H.~L.}\ \bibnamefont {Skriver}},\ }\bibfield
  {title} {\bibinfo {title} {Local kinetic-energy density of the airy gas},\
  }\href@noop {} {\bibfield  {journal} {\bibinfo  {journal} {Phys. Rev. A}\
  }\textbf {\bibinfo {volume} {61}},\ \bibinfo {pages} {052511(1)} (\bibinfo
  {year} {2000})}\BibitemShut {NoStop}%
\bibitem [{\citenamefont {Constantin}\ and\ \citenamefont
  {Ruzsinszky}(2009)}]{Constantin2009prb}%
  \BibitemOpen
  \bibfield  {author} {\bibinfo {author} {\bibfnamefont {L.~A.}\ \bibnamefont
  {Constantin}}\ and\ \bibinfo {author} {\bibfnamefont {A.}~\bibnamefont
  {Ruzsinszky}},\ }\bibfield  {title} {\bibinfo {title} {Kinetic energy density
  functionals from the airy gas with an application to the atomization kinetic
  energies of molecules},\ }\href@noop {} {\bibfield  {journal} {\bibinfo
  {journal} {Phys. Rev. B}\ }\textbf {\bibinfo {volume} {79}},\ \bibinfo
  {pages} {115117(7)} (\bibinfo {year} {2009})}\BibitemShut {NoStop}%
\bibitem [{\citenamefont {Kohn}\ and\ \citenamefont
  {Mattsson}(1998)}]{rf:kohn1998prl}%
  \BibitemOpen
  \bibfield  {author} {\bibinfo {author} {\bibfnamefont {W.}~\bibnamefont
  {Kohn}}\ and\ \bibinfo {author} {\bibfnamefont {A.~E.}\ \bibnamefont
  {Mattsson}},\ }\bibfield  {title} {\bibinfo {title} {Edge electron gas},\
  }\href@noop {} {\bibfield  {journal} {\bibinfo  {journal} {Phys. Rev. Lett.}\
  }\textbf {\bibinfo {volume} {81}},\ \bibinfo {pages} {3487} (\bibinfo {year}
  {1998})}\BibitemShut {NoStop}%
\bibitem [{\citenamefont {Wang}\ and\ \citenamefont
  {Teter}(1992)}]{rf:Wang1992prb}%
  \BibitemOpen
  \bibfield  {author} {\bibinfo {author} {\bibfnamefont {L.-W.}\ \bibnamefont
  {Wang}}\ and\ \bibinfo {author} {\bibfnamefont {M.~P.}\ \bibnamefont
  {Teter}},\ }\bibfield  {title} {\bibinfo {title} {Kinetic-energy functional
  of the electron density},\ }\href@noop {} {\bibfield  {journal} {\bibinfo
  {journal} {Phys. Rev. B}\ }\textbf {\bibinfo {volume} {45}},\ \bibinfo
  {pages} {13196} (\bibinfo {year} {1992})}\BibitemShut {NoStop}%
\bibitem [{\citenamefont {Wang}\ \emph {et~al.}(1998)\citenamefont {Wang},
  \citenamefont {Govind},\ and\ \citenamefont {Carter}}]{Wang1998prb}%
  \BibitemOpen
  \bibfield  {author} {\bibinfo {author} {\bibfnamefont {Y.~A.}\ \bibnamefont
  {Wang}}, \bibinfo {author} {\bibfnamefont {N.}~\bibnamefont {Govind}},\ and\
  \bibinfo {author} {\bibfnamefont {E.~A.}\ \bibnamefont {Carter}},\ }\bibfield
   {title} {\bibinfo {title} {Orbital-free kinetic-energy functionals for the
  nearly free electron gas},\ }\href
  {https://doi.org/10.1103/PhysRevB.58.13465} {\bibfield  {journal} {\bibinfo
  {journal} {Phys. Rev. B}\ }\textbf {\bibinfo {volume} {58}},\ \bibinfo
  {pages} {13465} (\bibinfo {year} {1998})}\BibitemShut {NoStop}%
\bibitem [{\citenamefont {Wang}\ \emph {et~al.}(1999)\citenamefont {Wang},
  \citenamefont {Govind},\ and\ \citenamefont {Carter}}]{Wang1999prb}%
  \BibitemOpen
  \bibfield  {author} {\bibinfo {author} {\bibfnamefont {Y.~A.}\ \bibnamefont
  {Wang}}, \bibinfo {author} {\bibfnamefont {N.}~\bibnamefont {Govind}},\ and\
  \bibinfo {author} {\bibfnamefont {E.~A.}\ \bibnamefont {Carter}},\ }\bibfield
   {title} {\bibinfo {title} {Orbital-free kinetic-energy density functionals
  with a density-dependent kernel},\ }\href@noop {} {\bibfield  {journal}
  {\bibinfo  {journal} {Phys. Rev. B}\ }\textbf {\bibinfo {volume} {60}},\
  \bibinfo {pages} {350} (\bibinfo {year} {1999})}\BibitemShut {NoStop}%
\bibitem [{\citenamefont {Huang}\ and\ \citenamefont
  {Carter}(2010)}]{Huang2010prb}%
  \BibitemOpen
  \bibfield  {author} {\bibinfo {author} {\bibfnamefont {C.}~\bibnamefont
  {Huang}}\ and\ \bibinfo {author} {\bibfnamefont {E.~A.}\ \bibnamefont
  {Carter}},\ }\bibfield  {title} {\bibinfo {title} {Nonlocal orbital-free
  kinetic energy density functional for semiconductors},\ }\href
  {https://doi.org/10.1103/PhysRevB.81.045206} {\bibfield  {journal} {\bibinfo
  {journal} {Phys. Rev. B}\ }\textbf {\bibinfo {volume} {81}},\ \bibinfo
  {pages} {1} (\bibinfo {year} {2010})}\BibitemShut {NoStop}%
\bibitem [{\citenamefont {Xia}\ \emph {et~al.}(2012)\citenamefont {Xia},
  \citenamefont {Huang}, \citenamefont {Shin},\ and\ \citenamefont
  {Carter}}]{Xia2012jcp}%
  \BibitemOpen
  \bibfield  {author} {\bibinfo {author} {\bibfnamefont {J.}~\bibnamefont
  {Xia}}, \bibinfo {author} {\bibfnamefont {C.}~\bibnamefont {Huang}}, \bibinfo
  {author} {\bibfnamefont {I.}~\bibnamefont {Shin}},\ and\ \bibinfo {author}
  {\bibfnamefont {E.~A.}\ \bibnamefont {Carter}},\ }\bibfield  {title}
  {\bibinfo {title} {Can orbital-free density functional theory simulate
  molecules?},\ }\href@noop {} {\bibfield  {journal} {\bibinfo  {journal} {J.
  Chem. Phys.}\ }\textbf {\bibinfo {volume} {136}},\ \bibinfo {pages}
  {084102(13)} (\bibinfo {year} {2012})}\BibitemShut {NoStop}%
\bibitem [{\citenamefont {Takahashi}(2022)}]{Takahashi2022ijqc}%
  \BibitemOpen
  \bibfield  {author} {\bibinfo {author} {\bibfnamefont {H.}~\bibnamefont
  {Takahashi}},\ }\bibfield  {title} {\bibinfo {title} {Development of kinetic
  energy density functional using response function defined on the energy
  coordinate},\ }\href@noop {} {\bibfield  {journal} {\bibinfo  {journal} {Int.
  J. Quantum Chem.}\ }\textbf {\bibinfo {volume} {122}},\ \bibinfo {pages}
  {e26969} (\bibinfo {year} {2022})}\BibitemShut {NoStop}%
\bibitem [{\citenamefont {Takahashi}(2018)}]{rf:Takahashi2018}%
  \BibitemOpen
  \bibfield  {author} {\bibinfo {author} {\bibfnamefont {H.}~\bibnamefont
  {Takahashi}},\ }\bibfield  {title} {\bibinfo {title} {Density-functional
  theory based on the electron distribution on the energy coordinate},\
  }\href@noop {} {\bibfield  {journal} {\bibinfo  {journal} {J. Phys. B:
  Atomic, Molecular and Optical Physics}\ }\textbf {\bibinfo {volume} {51}},\
  \bibinfo {pages} {055102(11pp)} (\bibinfo {year} {2018})}\BibitemShut
  {NoStop}%
\bibitem [{\citenamefont {Takahashi}(2020)}]{Takahashi2020jpb}%
  \BibitemOpen
  \bibfield  {author} {\bibinfo {author} {\bibfnamefont {H.}~\bibnamefont
  {Takahashi}},\ }\bibfield  {title} {\bibinfo {title} {Development of static
  correlation functional using electron distribution on the energy
  coordinate},\ }\href@noop {} {\bibfield  {journal} {\bibinfo  {journal} {J.
  Phys. B: At. Mol. Opt. Phys.}\ }\textbf {\bibinfo {volume} {53}},\ \bibinfo
  {pages} {245101(9pp)} (\bibinfo {year} {2020})}\BibitemShut {NoStop}%
\bibitem [{\citenamefont {Becke}(2003)}]{rf:becke2003jcp}%
  \BibitemOpen
  \bibfield  {author} {\bibinfo {author} {\bibfnamefont {A.~D.}\ \bibnamefont
  {Becke}},\ }\bibfield  {title} {\bibinfo {title} {A real-space model of
  nondynamical correlation},\ }\href@noop {} {\bibfield  {journal} {\bibinfo
  {journal} {J. Chem. Phys.}\ }\textbf {\bibinfo {volume} {119}},\ \bibinfo
  {pages} {2972} (\bibinfo {year} {2003})}\BibitemShut {NoStop}%
\bibitem [{\citenamefont {Harris}(1985)}]{rf:harris1985prb}%
  \BibitemOpen
  \bibfield  {author} {\bibinfo {author} {\bibfnamefont {J.}~\bibnamefont
  {Harris}},\ }\bibfield  {title} {\bibinfo {title} {Simplified method for
  calculating the energy of weakly interacting fragments},\ }\href@noop {}
  {\bibfield  {journal} {\bibinfo  {journal} {Phys. Rev. B}\ }\textbf {\bibinfo
  {volume} {31}},\ \bibinfo {pages} {1770} (\bibinfo {year}
  {1985})}\BibitemShut {NoStop}%
\bibitem [{\citenamefont {Wesolowsk}\ and\ \citenamefont
  {Warshel}(1993)}]{Wesolowski1993jpc}%
  \BibitemOpen
  \bibfield  {author} {\bibinfo {author} {\bibfnamefont {T.~A.}\ \bibnamefont
  {Wesolowsk}}\ and\ \bibinfo {author} {\bibfnamefont {A.}~\bibnamefont
  {Warshel}},\ }\bibfield  {title} {\bibinfo {title} {Frozen density functional
  approach for ab initio calculations of solvated molecules},\ }\href@noop {}
  {\bibfield  {journal} {\bibinfo  {journal} {J. Phys. Chem.}\ }\textbf
  {\bibinfo {volume} {97}},\ \bibinfo {pages} {8050} (\bibinfo {year}
  {1993})}\BibitemShut {NoStop}%
\bibitem [{\citenamefont {Elliott}\ \emph {et~al.}(2010)\citenamefont
  {Elliott}, \citenamefont {Burke}, \citenamefont {Cohen},\ and\ \citenamefont
  {Wasserman}}]{rf:elliott2010pra}%
  \BibitemOpen
  \bibfield  {author} {\bibinfo {author} {\bibfnamefont {P.}~\bibnamefont
  {Elliott}}, \bibinfo {author} {\bibfnamefont {K.}~\bibnamefont {Burke}},
  \bibinfo {author} {\bibfnamefont {M.~H.}\ \bibnamefont {Cohen}},\ and\
  \bibinfo {author} {\bibfnamefont {A.}~\bibnamefont {Wasserman}},\ }\bibfield
  {title} {\bibinfo {title} {Partition density-functional theory},\ }\href@noop
  {} {\bibfield  {journal} {\bibinfo  {journal} {Phys. Rev. A}\ }\textbf
  {\bibinfo {volume} {82}},\ \bibinfo {pages} {024501(1)} (\bibinfo {year}
  {2010})}\BibitemShut {NoStop}%
\bibitem [{\citenamefont {Perdew}\ \emph {et~al.}(1996)\citenamefont {Perdew},
  \citenamefont {Burke},\ and\ \citenamefont {Ernzerhof}}]{rf:perdew1996prl}%
  \BibitemOpen
  \bibfield  {author} {\bibinfo {author} {\bibfnamefont {J.~P.}\ \bibnamefont
  {Perdew}}, \bibinfo {author} {\bibfnamefont {K.}~\bibnamefont {Burke}},\ and\
  \bibinfo {author} {\bibfnamefont {M.}~\bibnamefont {Ernzerhof}},\ }\bibfield
  {title} {\bibinfo {title} {Generalized gradient approximation made simple},\
  }\href@noop {} {\bibfield  {journal} {\bibinfo  {journal} {Phys. Rev. Lett.}\
  }\textbf {\bibinfo {volume} {77}},\ \bibinfo {pages} {3865} (\bibinfo {year}
  {1996})}\BibitemShut {NoStop}%
\bibitem [{\citenamefont {Takahashi}\ \emph {et~al.}(2000)\citenamefont
  {Takahashi}, \citenamefont {Hori}, \citenamefont {Wakabayashi},\ and\
  \citenamefont {Nitta}}]{rf:takahashi2000cl}%
  \BibitemOpen
  \bibfield  {author} {\bibinfo {author} {\bibfnamefont {H.}~\bibnamefont
  {Takahashi}}, \bibinfo {author} {\bibfnamefont {T.}~\bibnamefont {Hori}},
  \bibinfo {author} {\bibfnamefont {T.}~\bibnamefont {Wakabayashi}},\ and\
  \bibinfo {author} {\bibfnamefont {T.}~\bibnamefont {Nitta}},\ }\bibfield
  {title} {\bibinfo {title} {A density functional study for hydrogen bond
  energy by employing real space grids},\ }\href@noop {} {\bibfield  {journal}
  {\bibinfo  {journal} {Chem. Lett.}\ }\textbf {\bibinfo {volume} {3}},\
  \bibinfo {pages} {222} (\bibinfo {year} {2000})}\BibitemShut {NoStop}%
\bibitem [{\citenamefont {Takahashi}\ \emph
  {et~al.}(2001{\natexlab{a}})\citenamefont {Takahashi}, \citenamefont {Hori},
  \citenamefont {Wakabayashi},\ and\ \citenamefont
  {Nitta}}]{takahashi2001jpca}%
  \BibitemOpen
  \bibfield  {author} {\bibinfo {author} {\bibfnamefont {H.}~\bibnamefont
  {Takahashi}}, \bibinfo {author} {\bibfnamefont {T.}~\bibnamefont {Hori}},
  \bibinfo {author} {\bibfnamefont {T.}~\bibnamefont {Wakabayashi}},\ and\
  \bibinfo {author} {\bibfnamefont {T.}~\bibnamefont {Nitta}},\ }\bibfield
  {title} {\bibinfo {title} {{Real Space Ab Initio Molecular Dynamics
  Simulations for the Reactions of OH Radical/OH Anion with Formaldehyde}},\
  }\href@noop {} {\bibfield  {journal} {\bibinfo  {journal} {J. Phys. Chem. A}\
  }\textbf {\bibinfo {volume} {105}},\ \bibinfo {pages} {4351} (\bibinfo {year}
  {2001}{\natexlab{a}})}\BibitemShut {NoStop}%
\bibitem [{\citenamefont {Takahashi}\ \emph
  {et~al.}(2001{\natexlab{b}})\citenamefont {Takahashi}, \citenamefont {Hori},
  \citenamefont {Hashimoto},\ and\ \citenamefont
  {Nitta}}]{rf:takahashi2001jcc}%
  \BibitemOpen
  \bibfield  {author} {\bibinfo {author} {\bibfnamefont {H.}~\bibnamefont
  {Takahashi}}, \bibinfo {author} {\bibfnamefont {T.}~\bibnamefont {Hori}},
  \bibinfo {author} {\bibfnamefont {H.}~\bibnamefont {Hashimoto}},\ and\
  \bibinfo {author} {\bibfnamefont {T.}~\bibnamefont {Nitta}},\ }\bibfield
  {title} {\bibinfo {title} {A hybrid qm/mm method employing real space grids
  for qm water in the tip4p water solvents},\ }\href@noop {} {\bibfield
  {journal} {\bibinfo  {journal} {J. Comp. Chem.}\ }\textbf {\bibinfo {volume}
  {22}},\ \bibinfo {pages} {1252} (\bibinfo {year}
  {2001}{\natexlab{b}})}\BibitemShut {NoStop}%
\bibitem [{\citenamefont {Takahashi}\ \emph {et~al.}(2003)\citenamefont
  {Takahashi}, \citenamefont {Hashimoto},\ and\ \citenamefont
  {Nitta}}]{rf:takahashi2003jcp}%
  \BibitemOpen
  \bibfield  {author} {\bibinfo {author} {\bibfnamefont {H.}~\bibnamefont
  {Takahashi}}, \bibinfo {author} {\bibfnamefont {H.}~\bibnamefont
  {Hashimoto}},\ and\ \bibinfo {author} {\bibfnamefont {T.}~\bibnamefont
  {Nitta}},\ }\bibfield  {title} {\bibinfo {title} {Quantum
  mechanical/molecular mechanical studies of a novel reaction catalyzed by
  proton transfer in ambient and supercritical states of water},\ }\href@noop
  {} {\bibfield  {journal} {\bibinfo  {journal} {J. Chem. Phys.}\ }\textbf
  {\bibinfo {volume} {119}},\ \bibinfo {pages} {7964} (\bibinfo {year}
  {2003})}\BibitemShut {NoStop}%
\bibitem [{\citenamefont {Chelikowsky}\ \emph
  {et~al.}(1994{\natexlab{a}})\citenamefont {Chelikowsky}, \citenamefont
  {Troullier},\ and\ \citenamefont {Saad}}]{rf:chelikowsky1994prl}%
  \BibitemOpen
  \bibfield  {author} {\bibinfo {author} {\bibfnamefont {J.~R.}\ \bibnamefont
  {Chelikowsky}}, \bibinfo {author} {\bibfnamefont {N.}~\bibnamefont
  {Troullier}},\ and\ \bibinfo {author} {\bibfnamefont {Y.}~\bibnamefont
  {Saad}},\ }\bibfield  {title} {\bibinfo {title} {Finite-difference
  pseudopotential method: electronic structure calculations without a basis},\
  }\href@noop {} {\bibfield  {journal} {\bibinfo  {journal} {Phys. Rev. Lett.}\
  }\textbf {\bibinfo {volume} {72}},\ \bibinfo {pages} {1240} (\bibinfo {year}
  {1994}{\natexlab{a}})}\BibitemShut {NoStop}%
\bibitem [{\citenamefont {Chelikowsky}\ \emph
  {et~al.}(1994{\natexlab{b}})\citenamefont {Chelikowsky}, \citenamefont
  {Troullier}, \citenamefont {Wu},\ and\ \citenamefont
  {Saad}}]{rf:chelikowsky1994prb}%
  \BibitemOpen
  \bibfield  {author} {\bibinfo {author} {\bibfnamefont {J.~R.}\ \bibnamefont
  {Chelikowsky}}, \bibinfo {author} {\bibfnamefont {N.}~\bibnamefont
  {Troullier}}, \bibinfo {author} {\bibfnamefont {K.}~\bibnamefont {Wu}},\ and\
  \bibinfo {author} {\bibfnamefont {Y.}~\bibnamefont {Saad}},\ }\bibfield
  {title} {\bibinfo {title} {High-order finite-difference pseudopotential
  method: an application to diatomic molecules},\ }\href@noop {} {\bibfield
  {journal} {\bibinfo  {journal} {Phys. Rev. B}\ }\textbf {\bibinfo {volume}
  {50}},\ \bibinfo {pages} {11355} (\bibinfo {year}
  {1994}{\natexlab{b}})}\BibitemShut {NoStop}%
\bibitem [{\citenamefont {Barnett}\ and\ \citenamefont
  {Landman}(1993)}]{Barnett1993prb}%
  \BibitemOpen
  \bibfield  {author} {\bibinfo {author} {\bibfnamefont {R.~N.}\ \bibnamefont
  {Barnett}}\ and\ \bibinfo {author} {\bibfnamefont {U.}~\bibnamefont
  {Landman}},\ }\bibfield  {title} {\bibinfo {title} {Born-oppenheimer
  molecular-dynamics simulations of finite systems: structure and dynamics of
  (h$_2$0)$_2$},\ }\href@noop {} {\bibfield  {journal} {\bibinfo  {journal} {J.
  Chem. Phys.}\ }\textbf {\bibinfo {volume} {48}},\ \bibinfo {pages} {2081}
  (\bibinfo {year} {1993})}\BibitemShut {NoStop}%
\bibitem [{\citenamefont {Becke}(1988)}]{rf:becke1988pra}%
  \BibitemOpen
  \bibfield  {author} {\bibinfo {author} {\bibfnamefont {A.~D.}\ \bibnamefont
  {Becke}},\ }\bibfield  {title} {\bibinfo {title} {{Density-functional
  exchange-energy approximation with correct asymptotic behavior}},\
  }\href@noop {} {\bibfield  {journal} {\bibinfo  {journal} {Phys. Rev. A}\
  }\textbf {\bibinfo {volume} {38}},\ \bibinfo {pages} {3098} (\bibinfo {year}
  {1988})}\BibitemShut {NoStop}%
\bibitem [{\citenamefont {Lee}\ \emph {et~al.}(1988)\citenamefont {Lee},
  \citenamefont {Yang},\ and\ \citenamefont {Parr}}]{rf:lee1988prb}%
  \BibitemOpen
  \bibfield  {author} {\bibinfo {author} {\bibfnamefont {C.}~\bibnamefont
  {Lee}}, \bibinfo {author} {\bibfnamefont {W.}~\bibnamefont {Yang}},\ and\
  \bibinfo {author} {\bibfnamefont {R.~G.}\ \bibnamefont {Parr}},\ }\bibfield
  {title} {\bibinfo {title} {{Development of the Colle-Salvetti
  correlation-energy formula into a functional of the electron density}},\
  }\href@noop {} {\bibfield  {journal} {\bibinfo  {journal} {Phys. Rev. B}\
  }\textbf {\bibinfo {volume} {37}},\ \bibinfo {pages} {785} (\bibinfo {year}
  {1988})}\BibitemShut {NoStop}%
\bibitem [{\citenamefont {Kleinman}\ and\ \citenamefont
  {Bylander}(1982)}]{rf:kleinman1982prl}%
  \BibitemOpen
  \bibfield  {author} {\bibinfo {author} {\bibfnamefont {L.}~\bibnamefont
  {Kleinman}}\ and\ \bibinfo {author} {\bibfnamefont {D.~M.}\ \bibnamefont
  {Bylander}},\ }\bibfield  {title} {\bibinfo {title} {Efficacious form for
  model pseudopotentials},\ }\href@noop {} {\bibfield  {journal} {\bibinfo
  {journal} {Phys. Rev. Lett.}\ }\textbf {\bibinfo {volume} {48}},\ \bibinfo
  {pages} {1425} (\bibinfo {year} {1982})}\BibitemShut {NoStop}%
\bibitem [{\citenamefont {Ono}\ and\ \citenamefont
  {Hirose}(1999)}]{rf:ono1999prl}%
  \BibitemOpen
  \bibfield  {author} {\bibinfo {author} {\bibfnamefont {T.}~\bibnamefont
  {Ono}}\ and\ \bibinfo {author} {\bibfnamefont {K.}~\bibnamefont {Hirose}},\
  }\bibfield  {title} {\bibinfo {title} {Timesaving double-grid method for
  real-space electronic-structure calculations},\ }\href@noop {} {\bibfield
  {journal} {\bibinfo  {journal} {Phys. Rev. Lett.}\ }\textbf {\bibinfo
  {volume} {82}},\ \bibinfo {pages} {5016} (\bibinfo {year}
  {1999})}\BibitemShut {NoStop}%
\bibitem [{\citenamefont {Bachelet}\ \emph {et~al.}(1982)\citenamefont
  {Bachelet}, \citenamefont {Hamann},\ and\ \citenamefont
  {Schl\"{u}ter}}]{Bachelet1982prb}%
  \BibitemOpen
  \bibfield  {author} {\bibinfo {author} {\bibfnamefont {G.~B.}\ \bibnamefont
  {Bachelet}}, \bibinfo {author} {\bibfnamefont {D.~R.}\ \bibnamefont
  {Hamann}},\ and\ \bibinfo {author} {\bibfnamefont {M.}~\bibnamefont
  {Schl\"{u}ter}},\ }\bibfield  {title} {\bibinfo {title} {Pseudopotentials
  that work: From h to pu},\ }\href@noop {} {\bibfield  {journal} {\bibinfo
  {journal} {Phys. Rev. B}\ }\textbf {\bibinfo {volume} {26}},\ \bibinfo
  {pages} {4199} (\bibinfo {year} {1982})}\BibitemShut {NoStop}%
\bibitem [{\citenamefont {Becke}\ and\ \citenamefont
  {Roussel}(1989)}]{rf:becke1989pra}%
  \BibitemOpen
  \bibfield  {author} {\bibinfo {author} {\bibfnamefont {A.~D.}\ \bibnamefont
  {Becke}}\ and\ \bibinfo {author} {\bibfnamefont {M.~R.}\ \bibnamefont
  {Roussel}},\ }\bibfield  {title} {\bibinfo {title} {Exchange holes in
  inhomogeneous system: a coordinate-space model},\ }\href@noop {} {\bibfield
  {journal} {\bibinfo  {journal} {Phys. Rev. A}\ }\textbf {\bibinfo {volume}
  {39}},\ \bibinfo {pages} {3761} (\bibinfo {year} {1989})}\BibitemShut
  {NoStop}%
\end{thebibliography}
\providecommand{\noopsort}[1]{}\providecommand{\singleletter}[1]{#1}%

\end{document}